\documentclass[useAMS,usenatbib,usegraphicx,nofootinbib]{mn2e}

\title[IDs and Redshifts of 1.1\,mm Sources in GOODS-N]{An AzTEC
  1.1\,mm survey of the GOODS-N field -- II. Multi-wavelength
  identifications and redshift distribution}

\author[Edward L. Chapin et al.]{ 
  \parbox[t]{\textwidth}{
    Edward~L.~Chapin$^1$\thanks{E-mail:~echapin@phas.ubc.ca},
    Alexandra~Pope$^2$\thanks{{\it Spitzer\/} Fellow},
    Douglas~Scott$^1$,
    Itziar~Aretxaga$^3$,
    Jason~E.~Austermann$^4$,
    Ranga-Ram~Chary$^5$,
    Kristen~Coppin$^6$,
    Mark~Halpern$^1$,
    David~H.~Hughes$^3$,
    James~D.~Lowenthal$^7$,
    Glenn~E.~Morrison$^{8,9}$,
    Thushara~A.~Perera$^{10}$,
    Kimberly~S.~Scott$^4$,
    Grant~W.~Wilson$^4$,
    Min~S.~Yun$^4$
  } 
  \\
  \\
  $^{1}$Dept. of Physics \& Astronomy, University of British Columbia,
  6224 Agricultural Road, Vancouver, B.C. V6T 1Z1, Canada\\
  $^{2}$National Optical Astronomy Observatory, 950 N. Cherry Ave.,
  Tucson, AZ 85719, USA \\
  $^{3}$Instituto Nacional de Astrof\'isica, \'Optica y Electr\'onica
  (INAOE), Aptdo. Postal 51 y 216, Puebla, Mexico\\
  $^{4}$Department of Astronomy, University of Massachusetts, Amherst,
  MA 01003, USA\\
  $^{5}$Division of Physics, Mathematics, and Astronomy, California
  Institute of Technology, Pasadena, CA 91125, USA \\
  $^{6}$Institute for Computational Cosmology, University of Durham,
  South Road, Durham DH1 3LE, UK \\
  $^{7}$Department of Astronomy, Smith College, Northampton, MA 01063, USA\\
  $^{8}$Institute for Astronomy, University of Hawaii, Honolulu, HI
  96822, USA \\
  $^{9}$Canada-France-Hawaii Telescope, Kamuela, HI 96743, USA \\
  $^{10}$Illinois Wesleyan University, P.O. Box 2900, Bloomington, IL
  61702-2900, USA}

\def\lsim{\mathrel{\lower2.5pt\vbox{\lineskip=0pt\baselineskip=0pt
           \hbox{$<$}\hbox{$\sim$}}}}

\def\gsim{\mathrel{\lower2.5pt\vbox{\lineskip=0pt\baselineskip=0pt
           \hbox{$>$}\hbox{$\sim$}}}}

\begin{document}

\label{firstpage}

\maketitle

\begin{abstract} 
  We present results from a multi-wavelength study of 29 sources
  (false detection probabilities $<5$\%) from a survey of the Great
  Observatories Origins Deep Survey-North field at 1.1\,mm using the
  AzTEC camera. Comparing with existing 850\,\micron\ SCUBA studies in
  the field, we examine differences in the source populations selected
  at the two wavelengths. The AzTEC observations uniformly cover the
  entire survey field to a 1-$\sigma$ depth of $\sim1$\,mJy. Searching
  deep 1.4\,GHz VLA, and {\it Spitzer\/} 3--24\,\micron\ catalogues,
  we identify robust counterparts for 21 1.1\,mm sources, and
  tentative associations for the remaining objects. The redshift
  distribution of AzTEC sources is inferred from available
  spectroscopic and photometric redshifts.  We find a median redshift
  of $z=2.7$, somewhat higher than $z=2.0$ for $850\,\mu$m-selected
  sources in the same field, and our lowest redshift identification
  lies at a spectroscopic redshift $z=1.1460$. We measure the
  850\,\micron\ to 1.1\,mm colour of our sources and do not find
  evidence for `850\,\micron\ dropouts', which can be explained by the
  low-SNR of the observations. We also combine these observed colours
  with spectroscopic redshifts to derive the range of dust
  temperatures $T$, and dust emissivity indices $\beta$ for the
  sample, concluding that existing estimates $T\sim30$\,K and
  $\beta\sim1.75$ are consistent with these new data.
\end{abstract}  


\section{Introduction}

Over the last decade observations at submillimetre and millimetre
wavelengths (350--1200\,\micron) have been used to detect a population
of luminous ($L_{\mathrm{IR}} = L(8$--1000\,\micron)$> 10^{12}$\,${\rm
  L}_\odot$) galaxies
\citep[e.g.][Austermann~et~al.~submitted]{smail1997,hughes1998,barger1998,eales1999,cowie2002,scott2002,borys2003,serjeant2003,webb2003,wang2004,greve2004,laurent2005,coppin2005,coppin2006,knudsen2006,bertoldi2007,khan2007,scott2008,greve2008,perera2008,greve2008,devlin2009}. These
objects, referred to as submillimetre galaxies (SMGs), are thought to
be high redshift analogues of the local Ultra Luminous Infra-red
Galaxy (ULIRG) population \citep{sanders1996}. Their large
luminosities and apparent lack of significant active galactic nuclei
(AGN) activity in most cases
\citep[e.g.][]{horn2000,bautz2000,almaini2003,alexander2005,pope2008}
imply star-formation rates $\gsim100$--1000\,${\rm
  M}_\odot$\,yr$^{-1}$. With orders of magnitude larger space density
at $z>1$ than in the present-day Universe, it is presently believed
that SMGs could represent an energetic early star-forming phase in the
process that produces giant elliptical galaxies, and a significant
fraction of the total star-formation rate density at $z \gsim 2$
\citep[see][for a review]{blain2002}.

The identification of multi-wavelength counterparts to SMGs is
hindered by the angular resolution of the current generation of
submillimetre (submm) instruments (typically $\sim10$--20\,arcsec),
and the high surface density and faintness of counterparts in the
optical/NIR, making unambiguous associations difficult. Significant
progress has been made in the field by first searching for candidates
in much lower surface density catalogues with higher astrometric
precision, in particular using 1.4\,GHz VLA interferometer maps, and
deep 24\,\micron\ {\it Spitzer\/} observations. This method works both
radio and mid-IR wavelengths, as in the submm, are biased toward the
detection of star-forming galaxies: the radio synchrotron emission has
a well-known correlation with the far-IR radiation that gets
redshifted in to the observed submm band, and 24\,\micron\ samples
primarily thermal emission from warmer dust in the vicinity of
star-forming regions. With the much improved positional uncertainties
of $\sim1$~arcsec offered by these radio and mid-IR data sets, it is
then possible to identify optical/NIR counterparts provided that they
are bright enough
\citep[e.g.][]{ivison2002,chapman2005,pope2006,ivison2007}.

In this paper we use this established procedure to identify
counterparts to SMGs detected in a 1.1\,mm map\footnote{Map available
  at {\texttt http://www.astro.umass.edu/AzTEC/}} of the Great
Observatories Origins Deep Survey-North \citep[GOODS-N,][]{perera2008}
using the Astronomical Thermal Emission Camera
\citep[AzTEC,][]{wilson2008}. GOODS-N is one of several well-studied
fields in the northern hemisphere that has the prerequisite radio
data, as well as deep {\sl Spitzer\/} coverage to identify
counterparts. There is also an impressive collection of optical
imaging ({\sl HST} and ground-based), and optical spectroscopy for
${>}\,1500$ targets with which to study the detailed properties of
individual objects once their positions are known.

Until recently, the most complete submm image towards GOODS-N was the
SCUBA 850\,\micron\ map of \citet{borys2003} \citep[see
also][]{pope2005,wall2008} that was produced from a heterogeneous
collection of data obtained by different groups with different
observing modes
\citep{hughes1998,barger2000,borys2002,serjeant2003,wang2004}. This
map produced a sample of nearly 40 sources, and was the subject of a
detailed multi-wavelength study
\citep{borys2004,pope2005,pope2006}. However, the spatially varying
noise of the SCUBA map, combined with the desire to search for even
higher-redshift sources that are expected to be more easily detected
at longer wavelengths due to the more favourable negative
$K$-correction \citep[e.g.][]{eales2003}, motivated the survey of
\citet{perera2008} to uniformly map the entire area at 1.1\,mm.  The
GOODS-N AzTEC map covers 245\,arcmin$^2$ (matching the {\em Spitzer}
coverage), has an 18~arcsec full-width half-maximum (FWHM) beam
(compared with 15~arcsec for SCUBA at 850\,\micron) and reaches a
uniform RMS depth of 0.96--1.16\,mJy\,beam$^{-1}$. Note that there is
also a map covering a similar area made using MAMBO at 1.2\,mm
\citep{greve2008}; those data have a smaller beam (11.1~arcsec FWHM),
but slightly less uniform coverage with noise varying between
0.7--1.2\,mJy\,beam$^{-1}$.

The 28 robust 1.1\,mm sources identified in \citet{perera2008} were
detected with significances $>3.8$-$\sigma$.  In this paper we present
potential counterparts for all of these sources, as well as one new
object that was obtained by deblending the brightest peak in the map,
$AzGN~1$, revealing a faint source that we label $AzGN~1.2$
\citep[corresponding to $GN~20$ and $GN~20.2$, respectively,
in][Pope~et.~al~submitted]{pope2006}. Note that the 1.1\,mm {\em
  deboosted} flux densities given in this paper in Table~\ref{tab:sed}
have been corrected for Eddington bias and in many cases have
signal-to-noise ratios (SNR) $<3$-$\sigma$: these values are the least
biased estimates for the true flux densities, but do not reflect the
robustness of the detections. The integrated negative tails of these
distributions were used to estimate false detection probabilities, and
a limit $p(S\,{<}\,0)<5$\% corresponds to the 3.8-$\sigma$ threshold
mentioned above. Using extensive simulations the actual spurious rate
for the entire sample was estimated to be 1 or 2 sources in
\citet{perera2008}.

We find robust counterparts for 21 objects, which we define to be
objects with false-identification probabilities $P<0.05$ within
6~arcsec. We also provide tentative identifications for the remaining
sources, considering counterparts up to 10~arcsec away and
$0.05<P<0.10$ (Section~\ref{sec:id}). These identifications enable us
to report radio--IR spectral energy distributions (SEDs) using the
VLA, SCUBA and {\em Spitzer\/} GOODS-N data.  For the robust list, we
identify spectroscopic redshifts for 7 objects in the literature, and
provide a combination of mid-IR and radio--(sub)mm photometric
redshifts for the remaining 13 sources (Section~\ref{sec:redshift}).
We compare our results with the existing SCUBA studies in this field
to: (i) identify differences in the redshift distributions of sources
selected at 850\,\micron\ and 1.1\,mm; (ii) evaluate the effectiveness
of searching for `850\,\micron\ dropouts' (objects detected at 1.1\,mm
but not at 850\,\micron) as a means for finding higher-redshift SMGs;
and (iii) probing the rest-frame distribution of dust properties of
SMGs consistent with measurements in the two bands
(Section~\ref{sec:sed}).

\section{Counterpart Identification}
\label{sec:id}

\begin{figure*}
\centering
\includegraphics[width=\linewidth]{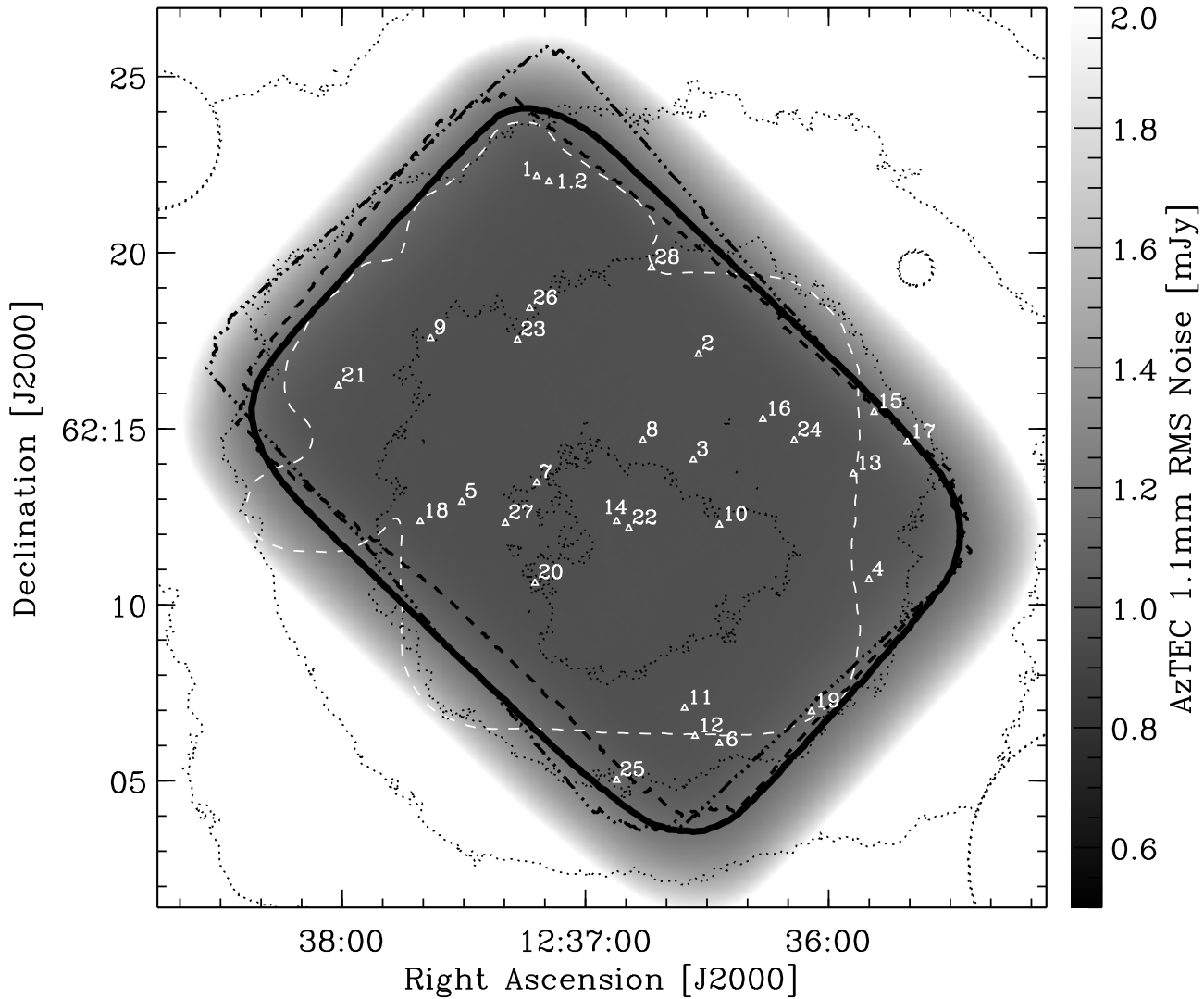}
\caption{The relative coverage of data sets in GOODS-N used in this
  paper. The greyscale indicates the RMS noise in the AzTEC 1.1\,mm
  map. The solid black contour corresponds to a noise of 1.16\,mJy in
  this map, and is the region within which AzTEC sources (white
  numbered triangles) were extracted. The white dashed contour
  indicates the SCUBA 850\,\micron\ coverage with a noise less than
  10\,mJy. The black dashed contour shows the MIPS 24\,\micron\
  coverage, and the dot-dashed lines the IRAC 3.6\,\micron\
  coverage. Finally, the thin dotted contours indicate the surface
  density of VLA 1.4\,GHz sources from the 4-$\sigma$ catalogue (0.5,
  1.0, 1.5 and 2.0\,arcmin$^{-2}$) measured in apertures with a radius
  of 0.1\,deg. For reference, the half-power radius of the radio map
  noise is about 0.25\,deg. The holes seen in these contours centred
  over 12:34:52 +62:03:41, 12:35:38 +62:19:32, and 12:38:48 +62:23:09
  are regions of the map that were excised due to sidelobe structure
  caused by especially bright sources.}
\label{fig:coverage}
\end{figure*}

\subsection{Radio and mid-IR matching catalogues}

The radio and {\em Spitzer\/} catalogues that we use to find
counterparts are generally the same as in \citet{pope2006}, and we
refer the reader to that paper for further details. The only
significant update to their analysis is an improved 1.4\,GHz VLA radio
map, with a 50\% reduction in the noise to $\sim4$--5\,$\mu$Jy RMS
across the AzTEC coverage region compared to that presented by
\citet{richards2000}, and about 25~per cent deeper than the map used
by \citet{pope2006}. The complete data set contains a total of
165.5\,hr of VLA 1.4\,GHz observations in A- (128.5\,hr), B- (28\,hr),
C- (\,7hr), and D- (2\,hr) configuration. These data were combined,
reduced, and imaged using AIPS. Full details of this analysis will be
presented in Morrison et al. (in preparation). While the final product
of that paper will be a 5-$\sigma$ catalogue, we have produced two
deeper catalogues for use in this work: (i) an approximately
4-$\sigma$ catalogue with a surface density of 1.80 arcmin$^{-2}$; and
(ii) a fainter 3-$\sigma$ radio catalogue with a surface density of
3.52\,arcmin$^{-2}$ that initially contains a much larger fraction of
spurious sources from which we select only objects that are coincident
within 1~arcsec of 3-$\sigma$ detections in the 24\,\micron\
catalogue. In both cases a radio catalogue of the given significance
is constructed using the AIPS source extraction task SAD in
signal-to-noise mode (using the uncorrected peak flux densities) which
uses an RMS map estimated from the task RMSD. After SAD detects the
sources over a region encompassing the AzTEC data, it then fits
Gaussian components to these sources, and in the process applies
corrections for both the bandwidth smearing and primary beam
attenuation. The final catalogues report these corrected flux
densities, including any spatially resolved structure that increases
the extent of the fitted Gaussians, without any additional `by-hand'
removal of problematic sources. The catalogues are therefore expected
to contain a number of false positives, and begin to suffer
incompleteness, at their respective flux density limits.  The
3-$\sigma$ catalogue covers a slightly smaller area, as it is limited
to the 24\,\micron\ coverage.  For several AzTEC sources around the
edge of the map we therefore use only the 4-$\sigma$ radio catalogue
to make identifications); see Fig~\ref{fig:coverage} for the relative
coverage of each data set.

\begin{figure}
\centering
\includegraphics[width=\linewidth]{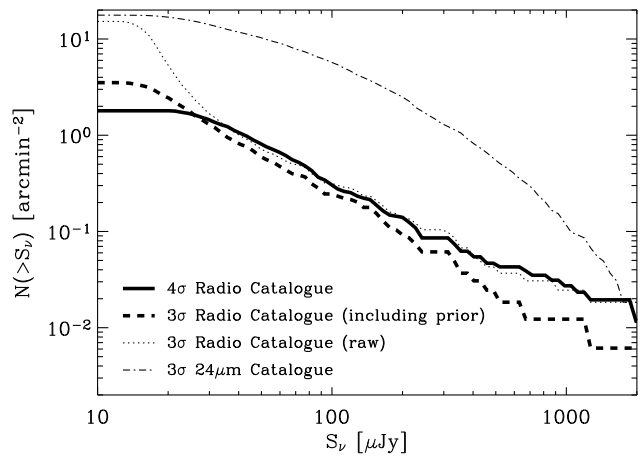}
\caption{Integral source counts in the radio and mid-IR matching
  catalogues. The primary catalogue is the 4-$\sigma$ radio catalogue
  and its integral source counts within the solid black contour in
  Fig.~\ref{fig:coverage} are shown by a solid line. A deeper
  3-$\sigma$ radio catalogue was produced by the intersection of a raw
  3-$\sigma$ radio catalogue with the 3-$\sigma$ 24\,\micron\
  catalogue. The counts from the two raw source catalogues are shown
  by the dotted and dot-dashed lines respectively, measured within the
  area common to the solid black and black dashed contours from
  Fig.~\ref{fig:coverage}. While the raw 3-$\sigma$ radio catalogue
  has a huge increase of sources at $S_{1.4} < 25\,\mu$Jy, indicating
  many spurious detections, the requirement of a 24\,\micron\
  counterpart rejects many of these sources, such that the remaining
  counts at fainter flux densities follow the trend established at the
  bright end, but extending to a fainter limit than the 4-$\sigma$
  catalogue.}
\label{fig:counts}
\end{figure}

In Fig.~\ref{fig:counts} the integral source counts for the radio and
mid-IR catalogues are shown. The horizontal axis has been extended to
sufficiently faint flux densities to show the point at which each
catalogue becomes incomplete (flat integral counts).  Clearly the raw
3-$\sigma$ radio catalogue contains many spurious detections at
$S_{1.4} \lsim 25\,\mu$Jy, as the counts diverge steeply from the
trend at brighter flux densities traced by both the 3-$\sigma$ and
4-$\sigma$ catalogues. However, the requirement of a 24\,\micron\
counterpart for each radio source drastically reduces the number of
candidate detections in this flux density regime, effectively
extending the faint counts to $\sim15$--20$\,\mu$Jy from the
$\sim20$--25\,$\mu$Jy achieved in the 4-$\sigma$ catalogue. We note
that while this technique enables us to extract fainter sources from
the radio map (rejecting many of the spurious noise peaks), the
requirement of 24\,\micron\ emission could lead to incompleteness
(i.e. real radio sources that are not detected with MIPS), as the
24\,\micron\ channel samples rest-frame spectral features
(e.g. polycyclic aromatic hydrocarbon emission and silicate
absorption) that are not directly related to the radio
emission. Nevertheless, it will be shown in the following sections
that even with these potentially incomplete radio catalogues we
identify a significant fraction of the 1.1\,mm sources.

\begin{figure}
\centering
\includegraphics[width=\linewidth]{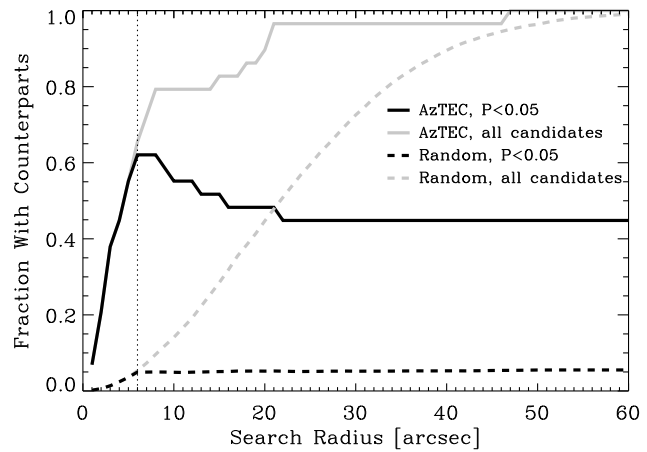}
\caption{The fraction of AzTEC sources with robust 4-$\sigma$ 1.4\,GHz
  ($P\,{<}\,0.05$) candidate identifications as a function of search
  radius (solid black line) compared with the fraction of sources with
  any candidate identifications (no cut on $P$, solid grey line).  For
  reference the black and grey dashed lines indicate the spurious
  counterpart detection rates using random positions with and without
  the $P\,{<}\,0.05$ cut, respectively. A search radius of 6~arcsec
  (vertical dotted line) is used since it gives the peak robust
  identification fraction.  It also happens to be where the random
  fraction corresponds to 5~per cent.  We chose this 6~arcsec radius
  to search for all candidates, even in the 3-$\sigma$ radio
  catalogue. In the event that no robust counterparts with $P<0.05$
  can be identified within 6~arcsec, the search radius is extended out
  to 10~arcsec to find tentative identifications.}
\label{fig:sr}
\end{figure}

For simplicity, a fixed search radius was used to identify potential
counterparts in the two catalogues, rather than a variable radius as a
function of AzTEC SNR \citep[see e.g.][]{ivison2007}. There is also a
relatively small dynamic range in AzTEC flux densities, and with
fairly constant noise, so that most of the positional uncertainties
would be similar in any case.  Given a search radius and list of
potential counterparts, the probability that a given candidate is a
random association,`$P$', is calculated following the prescription of
\citet{downes1986}, a method that is now used almost ubiquitously in
the submm literature. This technique accounts for the surface density
of sources in the matching catalogue as a function of brightness: if
two potential identifications of different brightnesses lie at the
same distance from the AzTEC source the rarer brighter object will be
assigned a lower value of $P$. We have used the raw measured integral
source counts in the matching catalogues (Fig.~\ref{fig:counts}) for
this calculation, rather than a model. Since the radio catalogues
undoubtedly contain a number of spurious sources near the detection
thresholds, this procedure will naturally account for them by
down-weighting the robustness of proposed faint identifications where
the surface density is much greater.

We also investigated the use of spatially-varying measurements of the
counts. Since the noise depth of the radio map falls off towards the
edge of the AzTEC coverage, so do the counts in the radio
catalogues. We chose an aperture with radius 0.1$^\circ$
(significantly smaller than the $\sim0.25^\circ$ half-power radius of
the VLA map noise) in which we measured the integral radio counts
centered over each position. For the 4-$\sigma$ radio catalogue this
resulted in a variation of the total surface density ranging from
about 2.4\,arcmin$^{-2}$ at the centre of the map, to
1.0--1.5\,arcmin$^{-2}$ along the edges (Fig.~\ref{fig:coverage}). We
experimented with measurements of $P$ using these modified counts at
the locations of each AzTEC source. With the increased source density
toward the centre of the map, values of $P$ are increased slightly,
and conversely values of $P$ are decreased slightly toward the
edges. However, the changes in individual values of $P$ are generally
$<20$\%, and the only effects of using this calculation on the final
list of robust identifications would be to add $AzGN~2$, as $P$ would
drop to 0.050 from 0.051 for the single radio source within 6\,arcsec,
and similarly, the second radio source near $AzGN~18$ would be added,
with $P$ dropping to 0.047 from 0.051 \citep[this object was already
identified as a radio double, $GN~38$, in][and the photometric
redshift estimates are consistent with them being at the same
distance]{pope2006}. Since the differences are small, and the simpler
calculation gives a slightly more conservative list of potential
counterparts, we elected to use only the radio and mid-IR number
counts averaged across the entire region of AzTEC coverage presented
in Fig.~\ref{fig:counts}. Average source counts were also used by
\citet{pope2006} and \citet{greve2008} in their calculations of $P$ to
find SMG counterparts in this field.

\subsection{Choice of search radius}
\label{sec:searchradius}

A search radius was chosen to provide a reasonable level of
completeness, while minimizing the number of false identifications. In
Fig.~\ref{fig:sr} we plot the fraction of AzTEC sources with at least
one such counterpart in the 4-$\sigma$ radio catalogue as a function
of search radius with $P<0.05$ (solid black line). The function
initially rises as most true counterparts are eventually detected,
reaches a peak of 62~per cent at 6--7~arcsec, and then drops as
potential identifications again become improbable due to the
increasing chance of a spurious detection with such a large search
radius. In contrast, if all radio sources within the search radius are
considered, the total fraction with potential counterparts continues
to grow (solid grey line). To demonstrate the effectiveness of cuts on
$P$ to lower the false identification rate, we repeat these
calculations using random positions. Using the cut on $P$ (dashed
black line) results in a plateau at the expected fraction of
$\sim0.05$, coincidentally, also at a search radius of 6~arcsec.
Without the cut, the spurious identification fraction (grey dashed
line) continues to rise to $\sim100$~per cent by a radius of
60~arcsec.  

The chosen search radius of 6~arcsec is smaller than those used for
SCUBA sources, typically in the range 7--8~arcsec for objects with
similar SNR
\citep[e.g.][]{ivison2002,webb2003,borys2004,pope2006,ivison2007},
despite AzTEC having a slightly larger beam. \citet{borys2004} used a
different method to estimate a search radius for counterparts to SCUBA
sources in GOODS-N, finding a value of 7~arcsec to be appropriate. We
repeated their analysis with our data, however, and concluded again
that we should use 6~arcsec. This test suggests that the smaller
resulting search radius is a property of the data, rather than the
method we used to calculate it. As a further consistency check, recent
SMA follow-up of AzTEC sources detected in other fields using the same
pointing model as that employed here has shown excellent
agreement. For example, \citet{younger2007} found positional
uncertainties $<4$~arcsec between SMA and AzTEC centroids of
millimetre sources in the COSMOS field detected with generally higher
SNR than the objects discussed in this paper.

Finally, we note in Fig.~\ref{fig:sr} that the fraction of AzTEC
sources with {\em any\/} 4-$\sigma$ radio candidates beyond 6~arcsec (no
cut on $P$, grey solid line) continues to grow significantly faster
(from $\sim65$\% to 80\%) than the spurious rate (grey dashed line,
from $\sim5$\% to 10\%) to a search radius of 8~arcsec.  This suggests
that $\sim70$\% of the entire AzTEC sample has counterparts in the
4-$\sigma$ radio catalogue, of which 10\%, about 3 objects, are within
6--8~arcsec of the AzTEC centroids. There is a radio source within
47~arcsec of every AzTEC centroid, although the additional objects
encountered beyond $\sim10$~arcsec are almost certainly chance
alignments.

\subsection{Radio and mid-IR identifications}

We first search for counterparts in the 4-$\sigma$ radio catalogue
with $P<0.05$, and if none are found within 6~arcsec we proceed to
search in the 3-$\sigma$ catalogue using the same radius and cut on
$P$. This procedure gives us a robust sample that will subsequently be
analyzed in detail. In cases where no such counterparts can be found,
we relax the search to identify more tentative counterparts out to a
radius of 10~arcsec, and/or $P<0.10$. Since these search parameters
are expected to result in a much higher fraction of chance alignments,
this extended catalogue is not used to measure properties of the
general 1.1\,mm galaxy population, and is simply included for
completeness (although statistically speaking most of these tentative
identifications are probably correct). In one case ($AzGN~27$) neither
radio catalogue yields a source within 6~arcsec, nor are there any
counterparts with $P<0.05$ out to 10~arcsec, so we search in the MIPS
24\,\micron\ catalogue (using the counts in Fig.~\ref{fig:counts} to
calculate $P$), finding two potential identifications (however, we are
still able to estimate radio flux densities by performing photometry
in the VLA map at the MIPS positions). While possibly containing the
correct counterparts for this particular object, the 24\,\micron\
catalogue is not, in general, as useful as the radio catalogues, due
to a significantly greater surface density and hence chance of
identifying random interlopers \citep[see also][]{ivison2007}.

\begin{figure*}
\centering
\includegraphics[width=\linewidth]{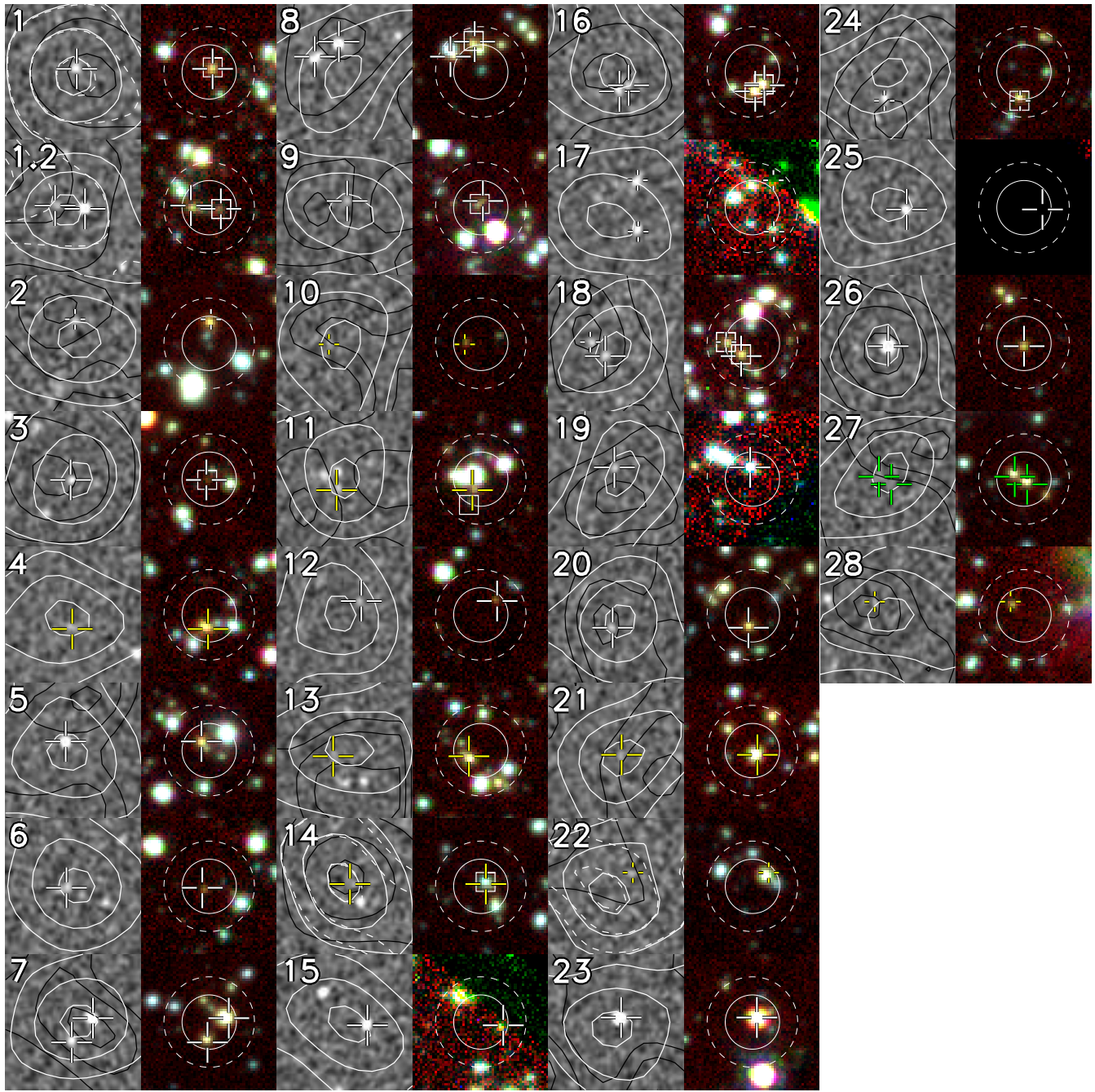}
\caption{30~arcsec$\times$30~arcsec postage stamps showing AzTEC
  counterpart identifications in GOODS-N. The left panels show the VLA
  1.4\,GHz map stretched between $-20\,\mu$Jy (black) and +30\,$\mu$Jy
  (white). The contours indicate fractions 0.1, 0.5 and 0.9 of the
  peak SNR in the AzTEC (white) and SCUBA (black; not always
  available) postage stamps. In addition, $AzGN~1$, $AzGN~1.2$,
  $AzGN~14$, and $AzGN~22$ have had the effects of nearby blended
  sources removed. The un-deblended AzTEC map contours are shown with
  dashed white lines for comparison.  Cross-hairs indicate the
  locations of potential counterparts from the catalogues in which
  they were originally identified: white for the 4-$\sigma$ radio
  catalogue; yellow for the 3-$\sigma$ radio catalogue; and green for
  the MIPS 24\,\micron\ catalogue. Long cross-hairs correspond to IDs
  with $P\le0.05$, and small cross-hairs for $P>0.05$. The right
  panels are false-colour images constructed from IRAC 3.6 (blue), 4.5
  (green) and 5.8\,\micron\ (red) exposures. Squares indicate the
  locations of radio/IR counterparts to SCUBA sources proposed in
  \citet{pope2006}. The solid and dashed circles indicate search radii
  of 6~arcsec and 10~arcsec, respectively. Approximately 1 or 2 of the
  AzTEC sources are expected to be false positives.}
\label{fig:thumbs}
\end{figure*}

In total this procedure yields at least one counterpart within
6~arcsec for 26/29 AzTEC sources, 22 of which are robust with
$P<0.05$. Of the robust identifications, 18 were found in the
4-$\sigma$ radio catalogue, 3 in the 3-$\sigma$ radio catalogue, and 1
in the MIPS 24\,\micron\ catalogue. Of the tentative associations
within 6~arcsec, 1 was identified in the 4-$\sigma$ radio catalogue,
and 3 in the 3-$\sigma$ radio catalogue. Finally, all of the remaining
3/29 AzTEC sources have at least one tentative counterpart in the
range $6 < r < 10$~arcsec in the 4-$\sigma$ radio catalogue. These
results are summarized in Fig.~\ref{fig:thumbs} and
Table~\ref{tab:id}. The spectral energy distributions for the proposed
counterparts are given in Table~\ref{tab:sed}. Additional notes for
each source can be found in Appendix~\ref{sec:notes}.

\subsection{False identification rate}

We can estimate the number of spurious identifications for the 26
potential counterparts found within 6~arcsec by summing their $P$
values, which gives 0.85. This implies that of those 26 AzTEC sources,
about one of the identifications within 6~arcsec is expected to be
spurious (noting that several sources have multiple proposed
identifications).  To understand how to interpret this result in terms
of overall completeness, we consider several factors.  First, the
AzTEC source list is expected to have $\sim1$--2 spurious detections
\citep{perera2008}.  Second, due to positional uncertainties, some of
the true counterparts will lie beyond 6~arcsec. We adopt the radial
offset distribution of \citet{ivison2007}, $r \exp (-r^2/2\sigma^2)$,
with $\sigma\sim0.6\times\mathrm{FWHM}/\mathrm{SNR}$, which assumes a
symmetric Gaussian beam and uncorrelated map noise. The cumulative
distribution of this analytic PDF results in a shape very similar to
the numerical simulations of \citet{scott2008} for AzTEC sources in
the COSMOS field.  Taking $\mathrm{FWHM}=18$~arcsec, and the SNR for
raw map flux densities (before deboosting), we would only expect to
encounter counterparts within 6~arcsec for 27.5/29 sources on average
if they were all real (neglecting positional uncertainties in the
matching catalogue). However, we would confidently expect to find {\it
  all\/} of the objects within 10~arcsec. Since we do not know which
(if any) of the AzTEC sources are false positives, we simply apply
this fraction to the expected number of real sources calculated above,
and find that there should be $\sim26$--27 real sources with true
positions within 6~arcsec of their 1.1\,mm centroids. This expectation
is consistent with our identification rate within 6~arcsec of
$\sim25$--26 out of 29 sources, although we stress that this
statistical argument does not necessarily imply that the unmatched
sources are spurious. We find $P$ values less than 0.05 for only 21 of
the sources encountered within 6~arcsec (excluding $AzGN~14$ as noted
below). While it may be the case that most of the remaining 5 sources
are in fact associated with the 1.1\,mm objects, it is also possible
that the true counterparts are simply fainter in the radio than the
catalogue limit. We cannot distinguish between these two cases in the
present study given the positional uncertainties in the AzTEC
centroids.

As a final warning, as with all studies that use $P$ to evaluate
chance alignment probabilities, there is an underlying assumption that
the matching catalogues are spatially unclustered. Two examples of
ways in which this condition could be broken are an increased surface
density of sources in the vicinity of SMGs due to multiple catalogue
entries being associated with the same physical structure (such as a
galaxy cluster), or foreground lensing of background objects \citep[an
effect which is in fact commonly used to identify faint SMGs,
e.g.][]{smail2002}. In both of these cases $P$ would be biased low. We
do not attempt to correct for these effects in this work, but we alert
the reader that evidence for such cases in GOODS-N will be discussed
in the following sections: the particularly complicated identification
of a counterpart for $AzGN~14$ \citep[also known
$HDF~850.1$,][]{hughes1998} which caused us to drop it from the
analysis in this paper; and the potential presence of a protocluster
at redshift $z\sim4$.

\subsection{Comparison with SCUBA identifications}
\label{sec:scubacompare}

Since 12 of the 29 objects discussed in this paper were also detected
by SCUBA (see Table~\ref{tab:id}) and identified in the radio and
mid-IR using similar techniques \citep{pope2006,wall2008}, it is
useful to compare proposed identifications to see how the new AzTEC
positions and deeper radio catalogues affect the results. We exclude
$AzGN~14$/$GN~14$ ($HDF~850.1$) from this comparison (and most of the
remaining analysis in this paper) as its true counterpart has been
under debate for some time due to the suspected obscuration by a
foreground elliptical \citep[see][and notes in
Appendix~\ref{sec:notes}]{dunlop2004,cowie2009}.  Of the remaining 11
overlapping sources, we propose identical counterparts for 8 of those
objects. For $AzGN~1.2$ we find that both the proposed counterpart of
\citet{pope2006} ($P=0.005$ in this paper) and a second fainter radio
object \citep[with $P=0.037$, also noted by][]{daddi2008} are both
robust identifications by our definition. In another similar case,
only one object from a radio double identified by \citet{pope2006} for
$AzGN~18$ is strictly a robust counterpart ($P=0.032$ in this paper),
while the second radio source misses the cut, with $P=0.051$. The only
object for which we propose a completely different counterpart is
$AzGN~11$/$GN~27$, which was classified as `tentative' in the SCUBA
map: the ACS/IRAC identification from \citet{pope2006} is absent in
the 1.4\,GHz map, and we instead propose a radio source that lies
slightly to the north, with $P=0.027$.

\section{Redshift Distribution}
\label{sec:redshift}

Some recent surveys at $1.1$--$1.2$\,mm claim to detect higher
redshifts than SCUBA surveys at 850\,\micron\
\citep[e.g.][]{younger2007,greve2008}, while others find redshift
distributions that are indistinguishable, possibly due to small
samples sizes \citep[e.g.][]{greve2004,bertoldi2007}.  For this AzTEC
survey we more accurately quantify any differences using greatly
improved redshift information, and comparing directly to the SCUBA
results in this field using the same methodology. While the
uncertainty in the 1.1\,mm distribution derived from our data is
large, due to the relatively small sample size, and cosmic variance
resulting from the area of GOODS-N, this differential measurement
yields a useful comparison between the two bands.

A number of groups have obtained spectroscopic redshifts in GOODS-N
\citep[e.g.][Stern et al.~in
preparation]{cohen2000,cowie2004,wirth2004,chapman2005,reddy2006,barger2008,
  pope2008, daddi2008, daddi2009}.  We found spectroscopic redshifts
for 10 of our proposed AzTEC counterparts in these publicly available
data-sets (see Table~\ref{tab:id}). However, one of those redshifts
($AzGN~8$) corresponds to the least favourable counterpart within the
search radius (see the discussion for this source in
Appendix~\ref{sec:notes}). Another similar case is $AzGN~27$ for which
a spectroscopic redshift has been obtained only for the more distant
of two potential counterparts. Finally, two radio sources that appear
to be associated with the single object $AzGN~7$ lie at redshifts
$z=1.996$ and $z=1.992$, and we assign a single redshift of $z=1.994$,
which is sufficiently precise for the purposes of this paper.
Therefore our sample of 21 sources with unambiguous identifications
contains only 7/21 sources with spectroscopic redshifts.  This
fraction is considerably lower than the 15/20 spectroscopic redshifts
for robust counterparts from the \citet{pope2006} SCUBA sample (also
excluding $HDF~850.1$), including the two new redshifts for $GN~20$
and $GN~20.2$ from \citet{daddi2008}, the redshift for $GN~10$ from
\citet{daddi2009} and two additional {\it Spitzer\/} IRS redshifts
from \citet{pope2008}. However, we are not surprised at this lower
rate since we rely on archival data for the redshifts of counterparts
to new AzTEC sources, whereas many of the spectroscopic redshifts for
SCUBA sources were obtained using targeted follow-up of proposed
identifications.

For sources without a spectroscopic redshift we first searched for
optical photometric redshift estimates. As none were found, we instead
employed two photometric redshift calculations using longer wavelength
data. The first, $z_{\mathrm{ir}}$, is a simple function of the {\it
  Spitzer\/} photometry with coefficients derived from fits to SCUBA
sources in GOODS-N with spectroscopic redshifts
\citep{pope2006}. Although this method does not assume any particular
SED, it benefits from the 1.6\,\micron\ stellar bump that produces a
strong characteristic feature in the observed IRAC 3.6--8.0\,\micron\
bands for sources at redshifts $1 \lsim z \lsim 4$
\citep{simpson1999,sawicki2002}. Such an empirical calculation may
provide less biased results than fitting spectral templates to the
data since there are degeneracies between the derived redshift and
assumptions about the starburst producing the stellar bump \citep[see
discussion in][]{yun2008}. While the residuals for this functional fit
are relatively small (with a maximum $\Delta z = 0.4$), no
uncertainties are provided in \citet{pope2006} for the remaining
sources. However, a similar photometric redshift estimator was derived
by \citet{wilson2008b}, and a comparison with spectroscopic redshifts
for SMGs from several fields (including SCUBA sources from both
GOODS-N and SHADES) finds that a 1-$\sigma$ error envelope $\Delta z =
0.15(1+z)$ is a reasonable uncertainty estimate for 15 SMGs at
redshifts $0 \lsim z \lsim 3$. We have compared the \citet{pope2006}
and \citet{wilson2008b} photometric redshift formulae for our data and
find that of the 7 robust identifications with spectroscopic redshifts
both methods provide estimates consistent with the spectroscopic
measurements for the 3 sources at $z<3$, within the
\citet{wilson2008b} error envelope. However, both estimates are biased
low at $z>3$, more so using the \citet{wilson2008b} redshift
estimator.  We also checked the scatter between the two methods for
all of the robust identifications finding that they both gave answers
compatible with the \citet{wilson2008b} uncertainty estimate. This
bias and scatter are unsurprising as both formulae were fit to SMGs
with spectroscopic redshifts $z\lsim 3$. In this work we assume the
1-$\sigma$ uncertainties are also $\Delta z_{\mathrm{ir}} =
0.15(1+z)$, but warn the reader that the redshifts of more distant
objects are probably systematically underestimated with this
technique.

The second photometric redshift indicator, $z_{\mathrm{rm}}$, uses the
radio and (sub)mm flux densities fit to templates of local galaxies
assuming the radio-IR correlation holds at high redshift
\citep[e.g.][]{caryun1999,aretxaga2007}. This method provides the only
redshift estimates for a handful of sources around the edges of the
AzTEC map where there is no {\it Spitzer\/} or optical coverage from
the GOODS survey (the entire AzTEC survey area overlaps with the
1.4\,GHz data).  Our redshifts are calculated using the same
methodology as \citet{aretxaga2007} and summarized in
Table~\ref{tab:id}. We note that the quoted 68 per cent confidence
intervals are theoretical estimations; \citet{aretxaga2007} checked
the scatter between photometric and spectroscopic redshifts for a
sample of SMGs with radio and submm data of similar quality to GOODS-N
finding an empirical symmetric 1-$\sigma$ scatter of $\Delta
z_{\mathrm{rm}} = 0.8$. We also adopt this uncertainty for consistency
with the empirically measured uncertainties for $z_{\mathrm{ir}}$.

\begin{figure}
\centering
\includegraphics[width=\linewidth]{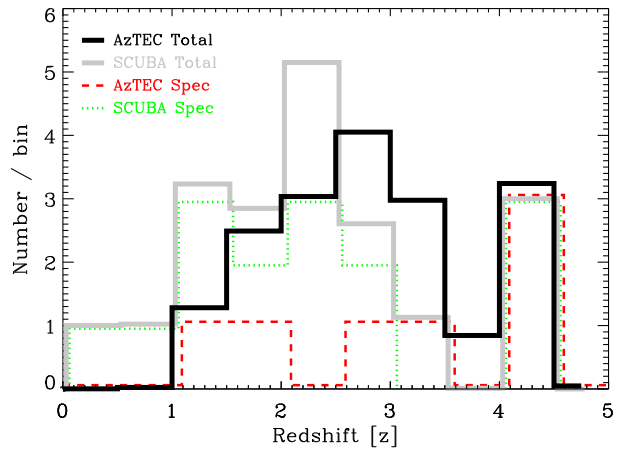}
\caption{ The redshift distribution of the 18 robustly identified
  1.1\,mm sources in GOODS-N (black histogram) with redshifts,
  adopting unique identifications from Table~\ref{tab:id} with the
  smallest $P$ values for each source. The grey histogram is the
  redshift distribution of 20 similarly robust SCUBA 850\,\micron\
  sources from \citet{pope2006}, excluding $GN~14$, and updating some
  spectroscopic redshifts based on \citet{daddi2008},
  \citet{daddi2009} and \citet{pope2008}. There is not an integral
  number of sources in each bin because the uncertainties in
  photometric redshifts have been included.  The results of K-S and
  M-W tests indicate chance probabilities $p_{\mathrm{KS}}=0.05$ and
  $p_{\mathrm{MW}}=0.01$, that both histograms were drawn from the
  same parent distribution (although they can be larger when
  individual redshift uncertainties are included, see
  Section~\ref{sec:redshift}). For comparison, the dashed red and
  green dotted histograms show the distribution of spectroscopic
  redshifts for AzTEC and SCUBA sources respectively. The excess
  number of objects in the $4.0<z<4.5$ bin may be members of a
  high-redshift protocluster \citep{daddi2008,daddi2009}. }
\label{fig:zhist}
\end{figure}

In Fig.~\ref{fig:zhist} we show the AzTEC redshift distribution of the
18/21 robust ($P<0.05$) identifications (black line) for which there
are redshifts (spectroscopic or photometric).  The remaining 3 objects
($AzGN~15$, $AzGN~19$, and $AzGN~25$) have only lower limits on their
redshifts, and have been excluded from the histogram and subsequent
analysis.  For the objects lacking spectroscopic redshifts we use the
photometric redshift estimates, including their uncertainty
distributions to divide them amongst several bins if necessary. When
both $z_{\mathrm{ir}}$ and $z_{\mathrm{rm}}$ are available, we
calculate the variance-weighted mean redshifts. We feel we are
justified in doing this because the two estimators depend on data and
spectral features at completely different wavelengths, and are
therefore independent of each other.  We then calculate the
distribution for 20 similarly robust 850\,\micron\ sources (grey line)
from \citet{pope2006} in the same way for comparison.  Note that, as
in this work, we have excluded $HDF~850.1$ ($GN~14$/$AzGN~14$) from
their redshift distribution. In addition we have updated redshifts for
several sources using new spectroscopic data
\citep{pope2008,daddi2008,daddi2009}. 

The median redshift of the AzTEC sample is $z=2.7$, with an
interquartile range of 2.1--3.4. In contrast, the median of the SCUBA
sample shown here is $z=2.0$ with an interquartile range 1.3--2.6. For
reference, the 14 SCUBA sources with spectroscopic redshifts (green
dotted histogram in Fig.~\ref{fig:zhist}) has a median $z=2.0$, and
the 7 AzTEC sources (red dashed histogram) $z=3.19$, both in good
agreement with the full distributions despite the small numbers of
objects. It is also worth noting that the spike of spectroscopic
redshifts in the $4.0<z<4.5$ bin seen at both wavelengths was found
via targeted follow-up to SCUBA sources by \citet{daddi2008} and
\citet{daddi2009} that are thought to be members of a proposed
protocluster at $z\sim4$. Although not included in the distribution, a
recent study by \citet{cowie2009} suggests that $HDF~850.1$ may also
be a member of this high-redshift structure.

We note that while the AzTEC sample appears to lie at slightly higher
redshift than the SCUBA sample, we have had to rely more heavily on
highly uncertain photometric estimates than in
\citet{pope2006}. However, the bias is likely to be toward lower
rather than higher redshifts due to the nature of $z_{\mathrm{ir}}$.

Next, taking the SCUBA and AzTEC redshift distributions at face value
(including photometric redshifts) we have used the Kolmogorov-Smirnov
(K-S) and Mann-Whitney $U$ (M-W) non-parametric tests for assessing
how different they are. These tests are fair since both samples were
drawn from the same region of space, and no extra uncertainty needs to
be included to account for cosmic variance.  Both methods operate on
discrete samples so we first assign the mean redshift to each object
from its uncertainty distribution. The K-S test, which is sensitive to
more general differences in the distributions (both the central values
and tails), gives a chance probability $p_{\mathrm{KS}} = 0.05$ that
both samples were drawn from the same parent redshift
distribution. The M-W test, which is mostly sensitive to differences
in the central values of the distributions, gives a smaller chance
probability $p_{\mathrm{MW}} = 0.01$. However, we note that the
uncertainties for objects with photometric redshifts can be as large
as $\Delta z \sim 0.8$, comparable to the width of the entire
population. To evaluate the spread in K-S and M-W probabilities that
are consistent with our sample, we generate 10,000 mock samples at
each wavelength drawing individual redshifts at random from the
uncertainty distributions for each object. We find that 68 per cent of
the time we obtain values $p_{\mathrm{KS}} < 0.15$ and
$p_{\mathrm{MW}} < 0.04$. These tests show that, even with large
individual uncertainties, the shift to higher redshifts at 1.1\,mm
compared to 850\,\micron\ appears to be statistically significant.

The SCUBA sample consists of a broader dynamic range in flux density
than the AzTEC sample, due to the varying map depths, and the fact
that different redshift populations may be present in the deep and
shallow regions of the map \citep{pope2006,wall2008}. It therefore may
be the case that at least some of the differences between these
distributions are a result of a depth rather than wavelength selection
effect.

One potential concern with this comparison is that, due to a bias to
higher redshifts at fainter flux densities
\citep[e.g.][]{chapman2005}, the deeper radio catalogues used for
matching in this survey simply detect more distant potential
counterparts than in \citet{pope2006}. We checked the distribution of
radio brightness with redshift for our sample and found that the 6
faintest proposed radio counterparts lie in the redshift interval
$2<z<3$. Removing them, while broadening the remaining redshift
distribution slightly, does not shift the median appreciably. However,
if we remove sources with even brighter radio flux densities, we in
fact begin to bias the sample to {\em higher} redshifts. Combined with
the fact that we find most of the same counterparts for sources that
appear in both the AzTEC and SCUBA surveys
(Section~\ref{sec:scubacompare}), we conclude that the intrinsic
rest-frame scatter of radio luminosities in SMGs dominates any
differences in the radio properties of 850\,\micron\ and 1.1\,mm
selected samples.

The {\it lowest\/} redshift that we find is $AzGN~23$ at $z=1.146$.
This demonstrates the ability of mm-wavelength surveys to effectively
select galaxies at $z>1$, with little contamination from nearby
objects.  Assuming that our identification procedure and redshift
estimates are correct, and given the completeness of our survey, there
is therefore little room for a significant tail to extremely high
redshifts. Since the negative $K$-correction at 1.1\,mm could in
principle enable us to detect SMGs easily out to a redshift $z\sim10$
\citep{blain2002}, the fact that objects at $z\gsim4.5$ do not appear
in our sample demonstrates that they do not exist in large quantities,
and would therefore require much larger surveys to find them. Only if
many of the identifications for these AzTEC sources are in fact more
complicated (as in the case of $HDF~850.1$) may the door still be open
for a significant fraction of the SMG population to lie at generally
higher redshifts ($z>4.5$).

\section{Spectral Energy Distributions}
\label{sec:sed}

With redshift estimates in hand, we are now in a position to probe the
rest-frame SEDs of our sample. Although we have photometry at a number
of wavelengths spanning 3.6\,\micron\ to $20\,$cm for most of the
objects, the most interesting new constraints that we place on these
SEDs is the shape of their rest-frame far-IR emission that peaks near
100\,\micron\ in the rest-frame, produced by thermal dust grain
emission. This emission accounts for most of the bolometric luminosity
in SMGs, and it is generally believed to be produced by
optically-obscured star formation in most cases
\citep[e.g.][]{blain2002}, much like locally observed ULIRGs. The
far-IR luminosity is therefore crucial for estimating star-formation
rates. The far-IR SED also provides a direct probe of the total dust
mass in a galaxy. However, both the bolometric luminosity and dust
mass are critically dependent on the dust temperature, $T$, and the
dust grain emissivity, $\beta$ \citep{hildebrand1983}. Due to a dearth
of data at the necessary wavelengths, spanning
$\sim100$--1000\,\micron, most authors either attempt to fit a simple
3-parameter modified blackbody spectrum for a population of dust
grains at a single temperature, $S_\nu = A\nu^\beta B_\nu(T)$ (where
$A$ is the amplitude), or adopt a single SED and normalize it to the
(sub)mm data point. Since only a single (sub)mm data point is usually
available, this latter compromise is often made. A census of recent
studies finds broad agreement that the most typical values are $T_{\rm
  d}=30$--$35\,$K for SMGs, with an allowed range that is somewhat
broader than this
\citep{chapman2005,kovacs2006,pope2006,huynh2007,coppin2008}.
However, the estimates of $T_{\rm d}$ and $\beta$ are highly
correlated, because of the limited range of wavelengths for which data
exist.

In GOODS-N the combination of 1.1\,mm and 850\,\micron\ flux densities
sample wavelengths longward of the rest-frame far-IR peak. The ratio
$S_{850}/S_{1.1}$ defines a family of 2-parameter SEDs ($T$ and
$\beta$) for each source which we will use to check for consistency
with previous measurements of the thermal SEDs of SMGs at typically
shorter wavelengths.

\subsection{Correcting for flux density bias}
We first estimate un-biased 1.1\,mm and 850\,\micron\ flux densities
for the AzTEC sources. As discussed in \citet{perera2008}, the 1.1\,mm
flux densities are biased high because they are selected from a low
SNR list of peaks coming from a counts distribution that falls steeply
with increasing brightness. The correction for this bias followed the
prescription of \citet{coppin2005}, and we adopt those posterior flux
density distributions here.  While this correction does not account
for the additional effect of source blending, the AzTEC GOODS-N survey
is shallower than the estimated confusion limit, and the sources that
appear to be confused have been fit explicitly in this paper using two
components. Rather than cross-matching the AzTEC catalogue with the
SCUBA catalogue to obtain 850\,\micron\ flux densities, which itself
suffers flux density bias (and since the SCUBA data are also too
shallow in some areas to provide flux densities for many of our
sources), we instead directly measure the 850\,\micron\ map at the
positions of proposed counterparts. Provided that these counterparts
are correct, and 850\,\micron\ source confusion is negligible, this
photometry yields un-biased 850\,\micron\ flux densities with
symmetric Gaussian uncertainties for {\em all\/} 24/29 AzTEC sources
that land within the region of SCUBA coverage. However, due to the
wide range in sensitivities only 9 objects have 850\,\micron\
detections with a significance of at least 3-$\sigma$.

\subsection{Searching for  `850\,\micron\ dropouts'}

It has been suggested that in regions where observations at both
850\,\micron\ and $\sim$1.1\,mm exist `850\,\micron\ dropouts',
i.e.~sources that are detected by AzTEC but not by SCUBA, can be used
to select predominantly higher-redshift sources
\citep{greve2004,greve2008}.  This technique is expected to work for
the same reason that the AzTEC redshift distribution is slightly
higher than the SCUBA sample: there is an increased submm negative
$K$-correction at 1.1\,mm compared to 850\,\micron\ \citep[i.e.~the
ratio of 850\,\micron\ to 1.2\,mm flux density, $S_{850}/S_{1.2}$,
decreases with redshift, seen for example in Figure~4
of][]{eales2003}.  In this study, we proceed by first testing the
hypothesis of a single intrinsic observed flux density ratio $R\equiv
S_{850}/S_{1.1}$, by measuring $R$ for several high-SNR objects
selected in the AzTEC map, and then searching for dropouts in the
SCUBA map relative to this average colour. We also repeat this
analysis in the opposite direction (for completeness), searching for
SCUBA sources that are dropouts in the AzTEC map.

We measure $R$ using sources with deboosted flux densities that have
significances $>3$-$\sigma$ in both bands: $AzGN~1$, $AzGN~3$,
$AzGN~7$ and $AzGN~8$, giving similar values 1.72, 1.96, 1.67 and 2.04
respectively. We adopt the mean $R=1.8$. For reference, thermal
emission from a galaxy with $T=30$\,K and $\beta=1.5$ at $z=2.5$ would
give an observed ratio 1.85. In order to compare our measurement with
values reported for SCUBA (850\,\micron) and MAMBO (1.2\,mm) overlap,
we use the same model SED to estimate how much larger the
$S_{850}/S_{1.2}$ ratio would be, finding that the scaled result is
2.3, near the centre of the distributions reported by
\citet{greve2004,greve2008}. Similarly we scale our result to estimate
the ratio $S_{890}/S_{1.1}$ for 890\,\micron\ SMA follow-up of AzTEC
sources, finding a ratio of 1.6. This value is consistent with
$1.4\pm0.3$ reported by \citet{younger2007}.

\begin{figure*}
\centering
\includegraphics[width=0.4\linewidth]{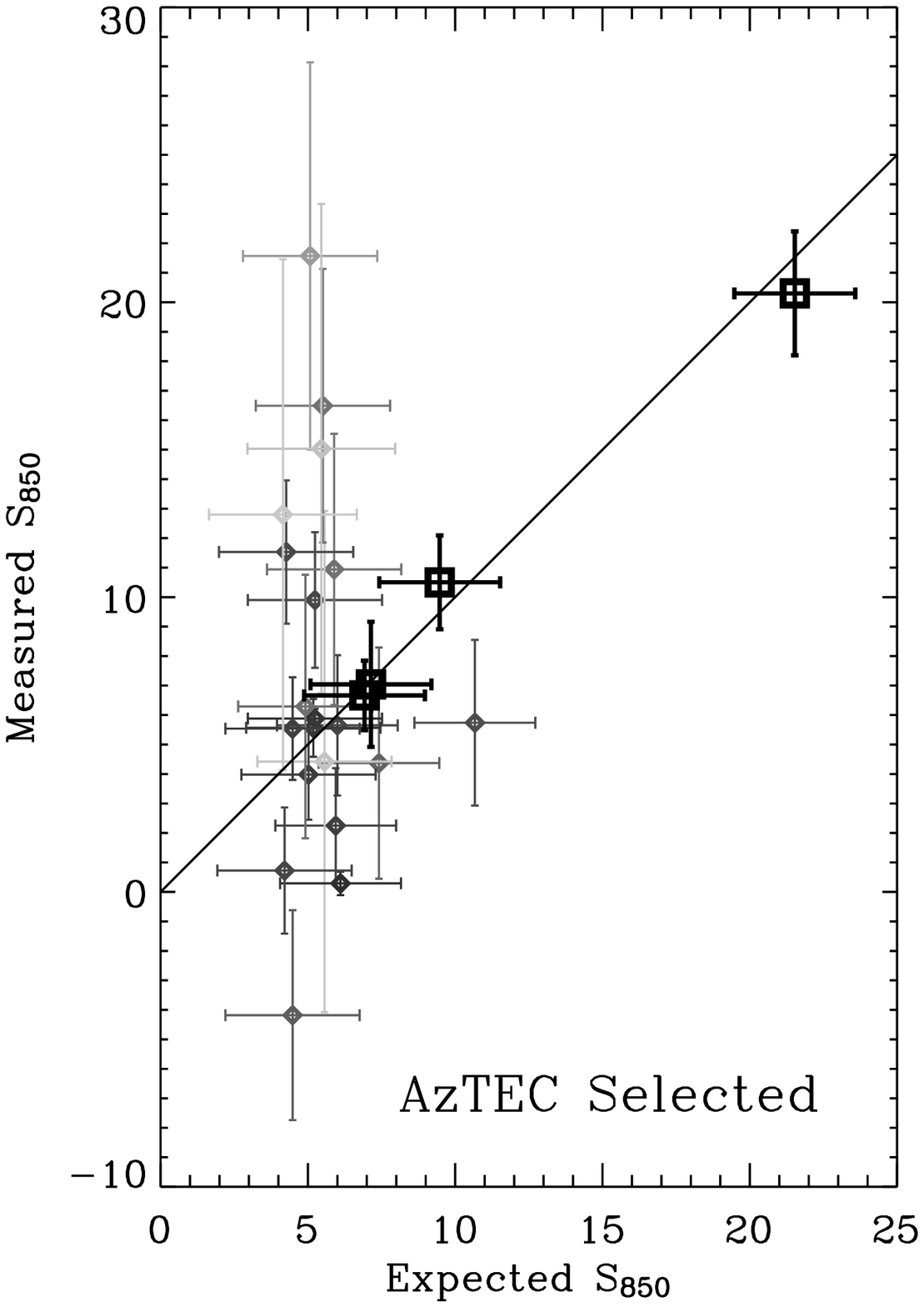} \hspace{0.05\linewidth}
\includegraphics[width=0.4\linewidth]{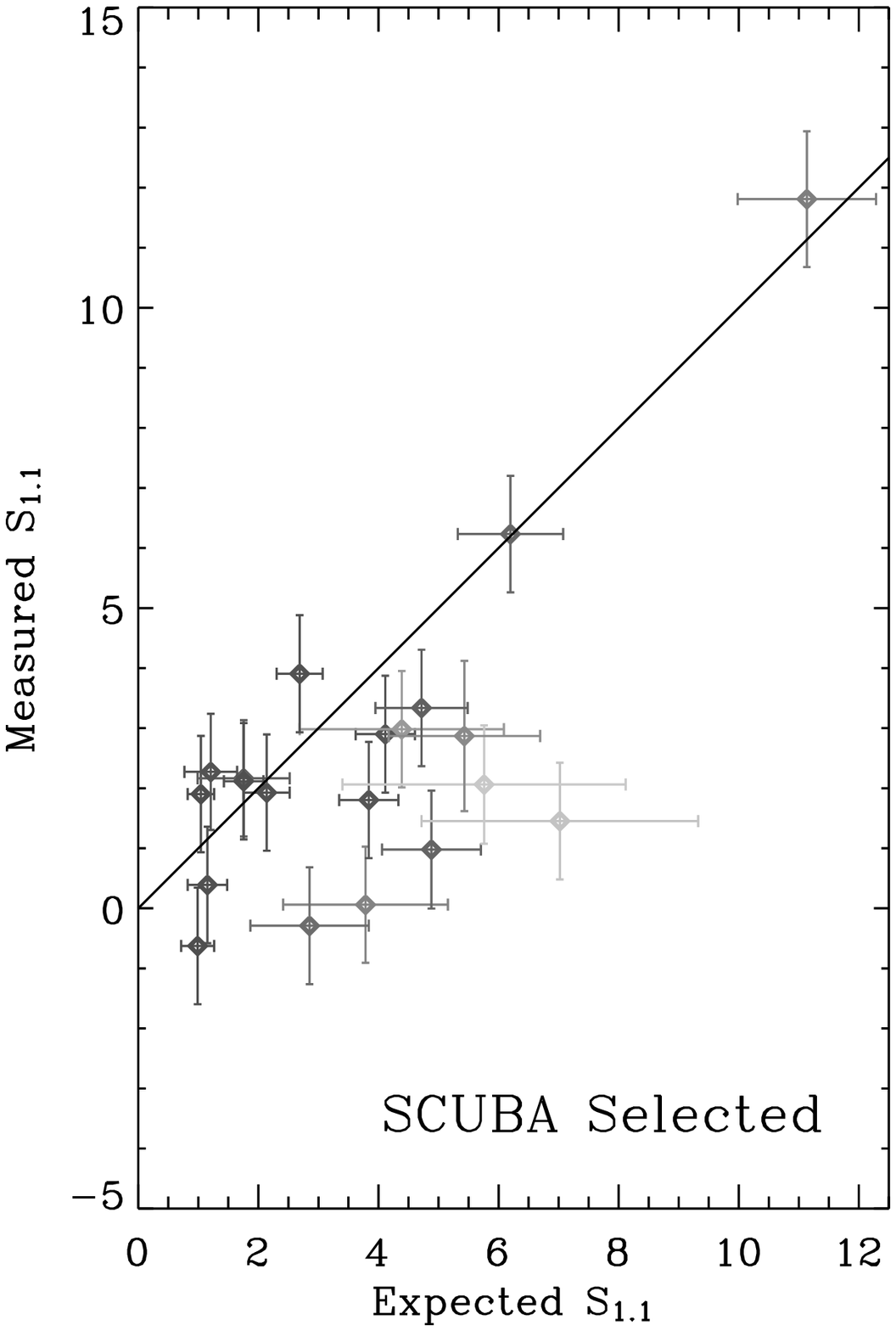}
\caption{Measured flux densities (vertical axes) compared with
  predicted flux densities (horizontal axes) for AzTEC (1.1\,mm) and
  SCUBA (850\,\micron) selected source lists.  {\bf left:} Measured
  SCUBA flux densities at the locations of counterparts proposed for
  AzTEC sources. The predicted SCUBA flux densities were scaled from
  the deboosted AzTEC measurements using the ratio of
  $S_{850}/S_{1.1}$ fit to the 4 most significant simultaneous
  detections in both the AzTEC and SCUBA maps (thick square symbols),
  yielding 1.8. We also plot the relation for the remaining 20 objects
  measured in both bands (diamonds). The symbols and error bars are
  coloured more lightly if they are less significant (combined
  observed and predicted 850\,\micron\ uncertainties). This plot shows
  symmetric scatter about the single-colour relation, and hence no
  clear evidence for 850\,\micron\ dropouts that would fall
  systematically below the line. {\bf right:} AzTEC flux densities at
  the locations of counterparts to SCUBA sources. The predicted values
  are derived from the deboosted SCUBA flux densities reported in
  \citet{pope2006} by dividing them by the same factor of 1.8. The low
  outliers are primarily low-SNR SCUBA measurements (larger horizontal
  error bars) suggesting that they have been insufficiently
  deboosted.}
\label{fig:ratios}
\end{figure*}

Next we use our measured $R$ to scale the deboosted 1.1\,mm flux
density distributions for the entire sample to 850\,\micron. For
simplicity we approximate the scaled 850\,\micron\ predicted flux
density distributions as Gaussians with mean values $s_{\rm m}$ given
by the modes, and standard deviations $\sigma_{\rm m}$ as half of the
68~per cent confidence intervals. If our hypothesis of a single
observed ratio were true, given the observed 850\,\micron\ data with
mean flux densities $s_{\rm d}$ and standard deviations $\sigma_{\rm
  d}$, we would expect the residuals $(s_{\rm m} - s_{\rm
  d})/\sqrt{\sigma_{\rm m}^2 + \sigma_{\rm d}^2}$ to be normally
distributed with mean 0 and standard deviation 1. For the 20 objects
that do not have 3-$\sigma$ detections at 850\micron\ we calculate the
sample mean and standard deviation of the residuals, giving $-0.1 \pm
1.4$. This calculation confirms that a ratio of 1.8 is a good estimate
for the central value of the observed distribution $S_{850}/S_{1.1}$
(left panel of Fig.~\ref{fig:ratios}).

Since the residual is broader than expected (by a factor
$\sim\sqrt{2}$), we conclude that the {\em intrinsic} spread in $R$
produces uncertainties of the same order as our measurement
errors. Note that the spread in $R$ for a $T=30\,K$, $\beta=1.5$ SED
from redshifts $z=1$--4 is only 2.1--1.6, so that part of the measured
spread must be due to a range of rest-frame dust emission spectra in
addition to the redshift distribution.  We note that this scatter is
roughly symmetric: the 850\,\micron\ measurements fall above the
expected values about as often as they fall below (in the left panel
of Fig.~\ref{fig:ratios}).

These calculations are biased to 1.1\,mm selected sources, and could
in principle be different for an 850\,\micron\ selected catalogue. We
therefore repeat the procedure, starting with deboosted flux densities
for the 20 robustly identified SCUBA sources mentioned in
Section~\ref{sec:redshift}, scaling them to 1.1\,mm by {\em dividing\/}
them by $R$, and comparing these predictions to photometry at the
locations of their counterparts in the AzTEC map (right panel of
Fig.~\ref{fig:ratios}). In this case the scatter is clearly
asymmetric, with a number of the SCUBA sources appearing fainter than
expected in the AzTEC map. However, these outliers are primarily
low-SNR SCUBA detections (indicated by the large horizontal error
bars), in particular $GN~3$, $GN~5$, $GN~7$, $GN~16$ and $GN~22$ which
have uncertainties ranging from 1.5--4.5\,mJy. We hypothesize that the
\citet{pope2006} deboosting recipe did not sufficiently correct these
sources, i.e. they should have been shifted further to the left in
this plot. This explanation is plausible because their deboosting
factors were extrapolated from those calculated for the SCUBA SHADES
survey \citep{coppin2006}, which were derived for sources with a
different noise distribution. Furthermore, the fact that the AzTEC
selected source list does not exhibit this problem (left panel of
Fig.~\ref{fig:ratios}) also points to an issue with the SCUBA
deboosting calculation, rather than the map itself.

\begin{figure*}
\centering
\includegraphics[width=0.49\linewidth]{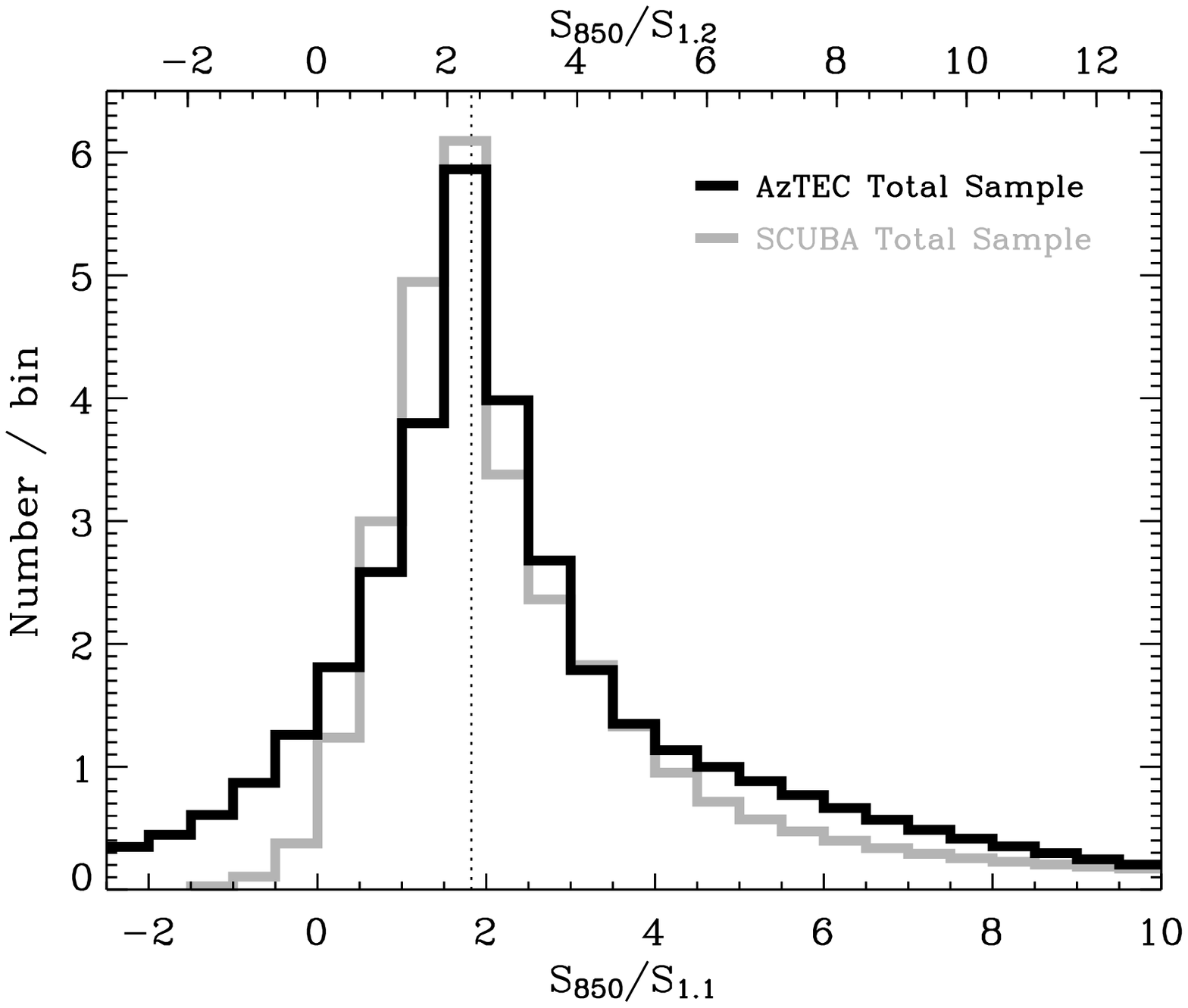}
\includegraphics[width=0.49\linewidth]{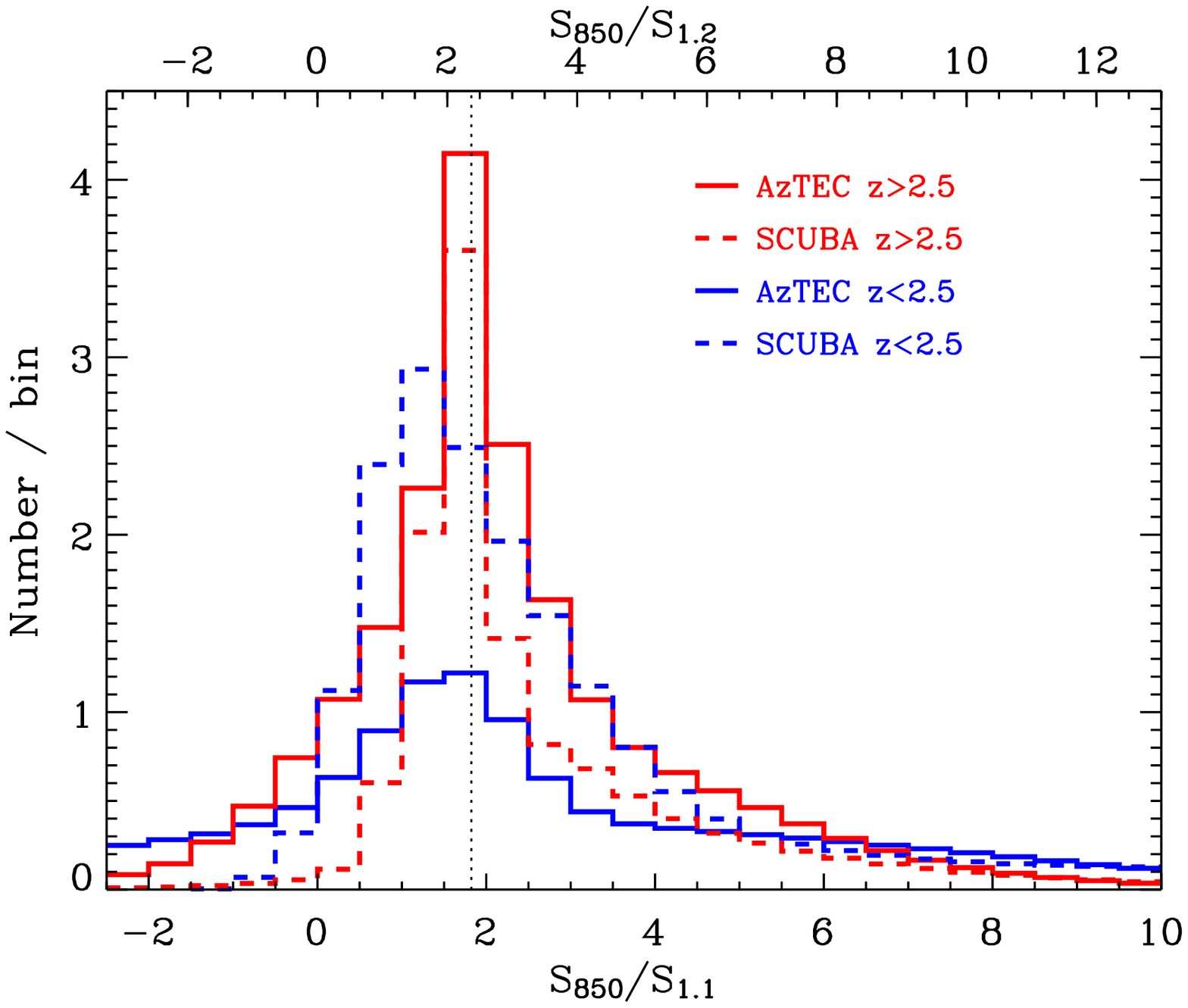}
\caption{Histograms of the $R=S_{850}/S_{1.1}$ flux ratio. In each
  case the vertical dotted line corresponds to a value of $R=1.8$
  calculated from high-SNR sources in Fig~\ref{fig:ratios}. The top
  horizontal axes assume a $\beta=1.5$, $T=30$\,K modified blackbody
  spectrum to predict the corresponding $S_{850}/S_{1.2}$ colours for
  comparison with MAMBO results.  {\bf left:\/} The black and grey
  histograms show the distributions of AzTEC and SCUBA selected
  sources, respectively. As in Fig.~\ref{fig:zhist} the uncertainty of
  each object is convolved with its error distribution before placing
  it in the histogram, such that higher-SNR measurements are more
  sharply peaked.  {\bf right:\/} Here the samples have been divided
  into subsets above and below redshift $z=2.5$. Although similar
  numbers of objects lie in each redshift bin, the higher redshift bin
  is dominated by the high-S/N detections of $AzGN~1$ and
  $AzGN~1.2$. }
\label{fig:colhist}
\end{figure*}

Finally, we use the AzTEC and SCUBA selected source lists to plot
histograms of the measured $S_{850}/S_{1.1}$ flux ratios in
Fig.~\ref{fig:colhist}. Here we have convolved each source with its
uncertainty distribution before adding it to the total histogram, so
that higher-SNR measurements are more sharply peaked and therefore
contribute more to the shape. In the left panel of this figure the
resulting histograms for each sample are shown to be nearly
indistinguishable, with modes that are coincident with the value
$R=1.8$ measured for the highest-SNR detections in the AzTEC map
(vertical dotted line). Since we also have redshift information, we
split the samples into objects above and below $z=2.5$, allowing us to
search for a systematic trend. For the AzTEC selected sample, there is
no significant difference in the ratios for low and high-redshift
objects. On the other hand, the mode of the SCUBA sample shows a mild
trend to {\em higher} values of $R$ with increasing redshift. However,
given the evidence for incorrect deboosting of the SCUBA flux
densities mentioned above we do not believe this trend is significant.

From these measurements we therefore conclude that there is no
evidence for 850\,\micron\ dropouts in the 1.1\,mm map. In fact, the
observed trend in the $S_{850}/S_{1.1}$ colour tends to go in the
opposite direction: AzTEC flux densities appear fainter than they
should be starting with a SCUBA selected list (but not vice
versa). How then can this result be reconciled with the measured shift
in the AzTEC redshift distribution compared to SCUBA
(Fig.~\ref{fig:zhist}), which can only be a result of enhanced
brightness at 1.1\,mm at larger values of $z$? Given the low-SNR
measurements used here, the wavelength proximity of the 850\,\micron\
and 1.1\,mm filters, and the fact that there appears to be a bias in
the SCUBA flux densities from \citet{pope2006}, we believe that trends
in the observed colour are completely obscured by uncertainties. Much
deeper (and higher spectral resolution) observations would be needed
to detect differences in the colours of 850\,\micron\ and 1.1\,mm
source populations.


\subsection{Constraints on $T$ and $\beta$}

Given the $T$--$\beta$ degeneracy, low SNR, and wavelength proximity
of the SCUBA 850\,\micron\ and AzTEC 1.1\,mm filters, the SED
constraints for individual objects are extremely noisy. However, with
this data-set it is possible to examine the constraints that we can
place on the {\em family\/} of $T$ and $\beta$ consistent with the
sample. Note that this is different to fitting SED templates based on
objects in the local Universe using all of the available radio--IR
photometry \citep[e.g.][]{pope2006,magnelli2009}. The purpose of such
fits is to give the best estimates of source properties, such as
bolometric luminosities and star-formation/AGN fraction. However, the
goal in this paper is to provide un-biased measurements that in the
future may be used to constrain redshift-dependent SED libraries.

\begin{figure}
\centering
\includegraphics[width=\linewidth]{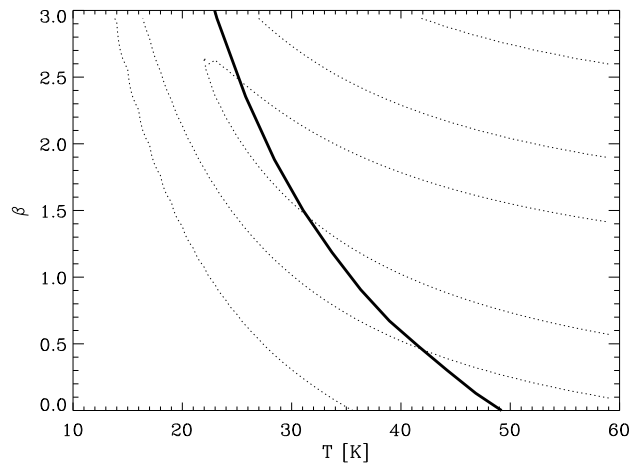}
\caption{Dotted lines are contours in the joint likelihood function
  (0.6, 0.1 and 0.001 times the maximum likelihood) demonstrating the
  family of values for $\beta$ and $T$ consistent with a simultaneous
  fit to the 5 galaxies for which spectroscopic redshifts have been
  obtained, and there exist at least 3-$\sigma$ measurements of the
  850\,\micron\ flux densities: $AzGN~1$, $AzGN~1.2$, $AzGN~3$,
  $AzGN~7$, and $AzGN~16$. The simple 3-parameter rest-frame SED $A
  \nu^\beta B_\nu(T)$ is redshifted, and compared to the observed
  850\,\micron\ and 1.1\,mm flux densities for each object with a grid
  of parameter values, and then marginalized over $A$. The final
  two-dimensional likelihood surface is the product of the five
  individual marginalized distributions. The solid line is the locus
  of SEDs that produce the same peak in $S(\nu)$ as the best-fit model
  from \citet{coppin2008}, $T=31$\,K and $\beta=1.5$ such that we can
  see the effect of an additional shorter-wavelength measurement to
  reject the highest-temperature / lowest-$\beta$ fits.}
\label{fig:chisq}
\end{figure}

We have extracted a subset of the galaxies for which there are robust
counterparts, 850\,\micron\ flux densities with a significance of at
least 3-$\sigma$, and spectroscopically measured redshifts: $AzGN~1$,
$AzGN~1.2$, $AzGN~7$, and $AzGN~16$. We then plot the joint likelihood
surface of the observed 850\,\micron\ and 1.1\,mm flux densities given
a range of models with $T$ and $\beta$ common to each object
(marginalizing over the amplitudes $A$ for each source as they are
irrelevant to the joint distribution) in Fig.~\ref{fig:chisq}. This
surface, as expected, produces a long anti-correlated valley between
$T$ and $\beta$.  For comparison we also plot the locus of SED models
that would produce the same peak $S(\nu)$ as the best-fit model from
\citet{coppin2008}, $T=31$\,K assuming $\beta=1.5$. As they use
shorter-wavelength 350\,\micron\ follow-up of SMGs, the constraints
are somewhat orthogonal to those described in this paper, showing the
ability of observations at those wavelengths to reject the
highest-temperature and lowest-$\beta$ SEDs.

There is some evidence that this AzTEC sample lies at slightly higher
redshift (Fig.~\ref{fig:zhist}), but the fact that we do not find
significant `850\,\micron\ dropouts' suggests that, within the
precision of the data, the SEDs are not different. At face value,
Fig.~\ref{fig:chisq} suggests that the best-fit value at the
intersection of the valley in the contour plot with the solid line
occurs at $T\sim30$\,K, and $\beta\sim1.75$. However, it is clear that
a large family of SEDs with temperatures ranging from $T\sim$25--40\,K
and corresponding $\beta\sim$2--1 are easily allowed.

\section{Conclusions}

We have used the rich multi-wavelength data-set in GOODS-N to identify
robust radio and IR counterparts for sources detected at 1.1\,mm using
AzTEC. Of the 29 sources, only 1 or 2 are expected to be false
positives. We find robust counterparts for 22 objects (false
identification probabilities $P<0.05$), although one object
($AzGN~14/HDF~850.1$) is dropped from the sample due to confusion
about its identification, and tentative associations for the remaining
8 objects are also provided.  These counterparts have an astrometric
precision of $\sim1$~arcsec, a significant improvement given the
18~arcsec FWHM AzTEC beam and low SNR.

We find spectroscopic redshifts for 7 of the robustly identified
sources in the literature, and provide photometric redshifts or limits
for the remaining objects. Restricting ourselves to the 18 objects
with robust counterparts and redshift estimates (spectroscopic or
photometric, excluding limits), we measure a median $z=2.7$, with an
interquartile range 2.1--3.4. The 850\,\micron\ sources in this
field, selected in a similar way, have a median redshift $z=2.0$ with
an interquartile range 1.3--2.6. We use K-S and M-W non-parametric
tests to evaluate the significance of this shift to higher redshift in
the 1.1\,mm map, finding chance probabilities $p_{\mathrm{KS}}=0.05$
and $p_{\mathrm{MW}}=0.01$ that both surveys sample the same redshift
population. Given the large uncertainties in individual photometric
redshifts, we used Monte Carlo simulations to evaluate the spread in
probabilities produced by the two tests consistent with our samples,
finding $p_{\mathrm{KS}} < 0.15$ and $p_{\mathrm{MW}} < 0.04$ at a
confidence level of 68 per cent.

For the entire overlapping region between the SCUBA and AzTEC maps, we
perform un-biased flux density measurements in the SCUBA map at
locations of identifications for AzTEC sources.  Using the 4 most
significant (3-$\sigma$) detections in the two maps we find a mean
observed 850\,\micron\ to 1.1\,mm flux density ratio $S_{850}/S_{1.1}
= 1.8$. For the remaining 20 sources, we observe a symmetric scatter
in the observed ratio which appears to be produced in equal quantities
by intrinsic spread in spectral properties, and measurement noise. We
also examine the ratios $S_{850}/S_{1.1}$ for objects selected at
850\,\micron, finding that they are also generally consistent with
this value, although it appears that the lower-SNR 850\,\micron\ flux
densities may be biased high. Finally, we unsuccessfully searched for
trends in this flux density ratio with redshift for both samples.  We
therefore do not see evidence for 850\,\micron\ dropouts in the
1.1\,mm map as reported in \citet{greve2004,greve2008}. While we
believe that such a trend must exist in the underlying SMG population
to produce the mild differences in the redshift distributions
mentioned above, it is undetectable when comparing $\sim$4-$\sigma$
detections in wavelength bands that are so close to eachother.

We test the hypothesis of a single temperature, $T$, and dust
emissivity index, $\beta$, for the ensemble of sources having robust
identifications and photometric redshift estimates. Given the
degeneracy between these parameters (since we have only two
photometric measurements at different wavelengths), we assume the same
mean rest-frame far-IR peak as found in other studies, finding that
$T=30$\,K and $\beta=1.75$ are consistent with all of the
data. However, given the SNR, these measurements still provide only a
weak constraint, and data at shorter rest-frame far-IR wavelengths
would be required to tighten up the allowable range of SEDs.  SCUBA-2
$450\,\mu$m, as well as SPIRE and new BLAST 250, 350\, and
500\,\micron\ surveys \citep[e.g.][]{devlin2009,dye2009} should be
particularly useful.

\section{Acknowledgements}

We thank D. Stern, E. MacDonald and M. Dickinson for providing their
unpublished Keck redshift for $AzGN~27$.  We also thank the anonymous
referee for their helpful comments. This research was supported by the
Natural Sciences and Engineering Research Council of Canada, and the
NSF grant AST05-40852.  AP acknowledges support provided by NASA
through the {\it Spitzer\/} Space Telescope Program, through a
contract issued by the Jet Propulsion Laboratory, California Institute
of Technology under a contract with NASA. IA and DHH acknowledge
partial support by CONACyT from research grants 39953-F and 39548-F.

\appendix

\section{Data Tables}
\label{sec:tables}

Here we provide proposed identifications and multi-wavelength
photometry for all of the sources. Table~\ref{tab:newflux} provides
updated positions and raw map flux densities for AzTEC 1.1\,mm sources
that required deblending. Tables~\ref{tab:id} and \ref{tab:sed}
summarize the identifications and SEDs respectively.

\begin{table}
  \centering
  \caption{New source positions and raw 1.1\,mm map flux densities 
    resulting from simultaneous two-source fits. Source Az~1.2 is a new
    object in this paper, whereas the other three were originally detected
    in AzTEC maps in \citet{perera2008}.}
  \begin{tabular}{lccc}
    \hline
    AzTEC & R.A. &Dec. & Map Flux Density \\
  ID &  ($^{\mathrm h}$\ \ \ $^{\mathrm m}$\ \ \ $^{\mathrm s}$) & 
  ($^{\circ}$\ \ \ $^{\prime}$\ \ \ $^{\prime\prime}$) & (mJy) \\

  \hline
  1   & 12 37 11.99 & +62 22 11.1 & $12.73 \pm 0.99$ \\
  1.2 & 12 37 09.15 & +62 22 02.1 & $ 4.14 \pm 0.98$ \\
  14  & 12 36 52.23 & +62 12 25.2 & $ 4.11 \pm 0.97$ \\
  22  & 12 36 49.12 & +62 12 13.1 & $ 4.56 \pm 0.97$ \\
  \hline
\end{tabular}
\label{tab:newflux}
\end{table}

\begin{table*}
  \caption{Radio and {\it Spitzer\/} identifications of AzTEC sources 
    (procedure described in Section~\ref{sec:id}). 
    Counterpart distances in brackets employed a 10~arcsec search 
    radius. $P$ values in boldface emphasize robust counterparts with values 
    $<0.05$.  Spectroscopic redshifts are given in the column labelled 
    $z_{\mathrm{spec}}$ (references for these measurements given in 
    Appendix~\ref{sec:notes}). Photometric redshifts based
    on {\it Spitzer\/} IR flux densities from Table~\ref{tab:sed} are calculated
    using Equation~2 from \citet{pope2006} and given in the penultimate column, 
    $z_{\mathrm{ir}}$. These redshifts have uncertainties 
    $\Delta z_{\mathrm{ir}} = 0.15(1+z)$, and are biased low at $z>3$. 
    Photometric redshifts
    based on the (sub)mm-to-radio colours are given in the last column, 
    $z_{\mathrm{rm}}$. The quoted 68 per cent uncertainties are theoretically 
    derived; in this paper we assume an empirically measured symmetric error 
    $\Delta z_{\mathrm{rm}} = 0.8$.}
  \begin{tabular}{llcclccllccc}
    \hline
    AzTEC & SCUBA & \multicolumn{3}{c}{Radio} & \multicolumn{3}{c}{Spitzer} & & 
    \multicolumn{3}{c}{Redshift} \\

    ID & ID & R.A. & Dec. & Dist. & R.A. & Dec.  & Dist. & $P$ & 
    $z_{\mathrm{spec}}$ & $z_{\mathrm{ir}}$ & $z_{\mathrm{rm}}$ \\

    & & ($^{\mathrm h}$\ \ \ $^{\mathrm m}$\ \ \ $^{\mathrm s}$) &  
    ($^{\circ}$\ \ \ $^{\prime}$\ \ \ $^{\prime\prime}$) & ($^{\prime\prime}$) &
    ($^{\mathrm h}$\ \ \ $^{\mathrm m}$\ \ \ $^{\mathrm s}$) &  
    ($^{\circ}$\ \ \ $^{\prime}$\ \ \ $^{\prime\prime}$) & ($^{\prime\prime}$) &&&&\\ 
    
  \hline

1 & 20 &  12 37 11.88 &   +62 22 11.8 & 1.0 &  12 37 11.88 &   +62 22 12.1 & 1.3 & {\bf 0.003} & {\bf 4.055} & 2.7 & $3.8^{+1.2}_{-0.7}$ \\
1.2 & 20.2 &  12 37 08.78 &   +62 22 01.8 & 2.6 &  12 37 08.77 &   +62 22 01.8 & 2.7 & {\bf 0.005} & {\bf 4.052} & 2.5 & $3.1^{+1.2}_{-0.2}$ \\
 &  &  12 37 09.73 &   +62 22 02.5 & 4.1 &  12 37 09.57 &   +62 22 02.1 & 2.9 & {\bf 0.037} & \ldots & 3.1 & $2.5^{+0.8}_{-0.8}$ \\
2 & \ldots &  12 36 31.93 &   +62 17 14.7 & 5.3 &  12 36 31.92 &   +62 17 14.6 & 5.2 & 0.051 & \ldots & 3.2 & $2.4^{+2.3}_{-0.1}$ \\
3 & 10 &  12 36 33.42 &   +62 14 08.7 & 0.6 &  12 36 33.40 &   +62 14 08.4 & 0.6 & {\bf 0.002} & {\bf 4.042} & 2.3 & $3.1^{+1.4}_{-0.2}$ \\
4 & \ldots &  12 35 50.26 &   +62 10 41.3 & 3.1 &  12 35 50.35 &   +62 10 41.8 & 2.7 & {\bf 0.030}$^a$ & \ldots & 2.9 & $2.4^{+2.2}_{-0.2}$ \\
5 & \ldots &  12 37 30.78 &   +62 12 58.7 & 2.6 &  12 37 30.75 &   +62 12 58.4 & 2.2 & {\bf 0.007} & \ldots & 2.0 & $2.1^{+1.6}_{-0.9}$ \\
6 & \ldots &  12 36 27.26 &   +62 06 05.7 & 1.6 &  12 36 27.21 &   +62 06 05.7 & 1.2 & {\bf 0.007} & \ldots & 3.0 & $2.9^{+1.4}_{-1.1}$ \\
7 & 39 &  12 37 11.32 &   +62 13 30.9 & 4.4 &  12 37 11.34 &   +62 13 31.0 & 4.3 & {\bf 0.014} & {\bf 1.996} & 1.7 & $2.7^{+0.9}_{-0.8}$ \\
 &  &  12 37 11.99 &   +62 13 25.6 & 4.5 &  12 37 11.99 &   +62 13 25.7 & 4.4 & {\bf 0.032} & {\bf 1.992} & 2.0 & $>$2.3 \\
8 & 12 &  12 36 46.04 &   +62 14 48.6 & (6.9) &  12 36 46.07 &   +62 14 48.8 & (7.0) & {\bf 0.037} & \ldots & 2.0 & $3.0^{+0.6}_{-1.0}$ \\
 &  &  12 36 46.80 &   +62 14 45.3 & (7.5) &  12 36 46.88 &   +62 14 47.2 & (9.0) & {\bf 0.050} & {\bf 2.006} & 0.7 & $2.9^{+0.7}_{-1.1}$ \\
9 & 37 &  12 37 38.16 &   +62 17 37.0 & 1.6 &  12 37 38.26 &   +62 17 36.4 & 0.9 & {\bf 0.013} & {\bf 3.1900} & 2.4 & $>$3.0 \\
10 & \ldots &  12 36 27.54 &   +62 12 17.8 & 3.5 &  12 36 27.48 &   +62 12 18.0 & 3.1 & 0.066$^a$ & \ldots & 3.0 & $2.1^{+2.2}_{-0.7}$ \\
11 & 27 &  12 36 35.89 &   +62 07 03.8 & 3.1 &  &  &  & {\bf 0.027} & \ldots & \ldots & $2.8^{+1.3}_{-0.2}$ \\
12 & \ldots &  12 36 32.65 &   +62 06 21.1 & 4.7 &  12 36 32.65 &   +62 06 21.3 & 4.9 & {\bf 0.047} & \ldots & 2.8 & $>$1.3 \\
13 & \ldots &  12 35 54.23 &   +62 13 43.8 & 2.9 &  12 35 54.28 &   +62 13 43.4 & 3.3 & {\bf 0.033}$^a$ & \ldots & 2.2 & $2.6^{+0.9}_{-1.5}$$^c$ \\
14 & 14 &  12 36 52.07$^d$ &   +62 12 25.7$^d$ & 1.2 &  &  &  & {\bf } & \ldots & \ldots & \ldots \\
15 & \ldots &  12 35 47.93 &   +62 15 29.2 & 5.0 &  12 35 48.09 &   +62 15 29.3 & 3.9 & {\bf 0.018} & \ldots & \ldots & $>$1.6 \\
 &  &  &  &  &  12 35 47.87 &   +62 15 28.2 & 5.7 & {\bf } & \ldots & \ldots & $>$1.6 \\
16 & 04 &  12 36 16.09 &   +62 15 13.8 & 4.3 &  12 36 16.10 &   +62 15 13.6 & 4.5 & {\bf 0.038} & {\bf 2.578} & 2.0 & $2.4^{+2.1}_{-0.2}$ \\
 &  &  12 36 15.80 &   +62 15 15.1 & 4.0 &  12 36 15.82 &   +62 15 15.4 & 3.6 & {\bf 0.039} & \ldots & 4.5 & $3.0^{+0.3}_{-0.2}$ \\
17 & \ldots &  12 35 39.92 &   +62 14 42.1 & (7.6) &  12 35 39.95 &   +62 14 40.8 & (6.5) & 0.062 & \ldots & \ldots & $>$1.7 \\
 &  &  12 35 39.91 &   +62 14 30.8 & (7.1) &  12 35 39.97 &   +62 14 30.7 & (6.9) & 0.067 & \ldots & \ldots & $>$1.7 \\
 &  &  &  &  &  12 35 39.76 &   +62 14 30.7 & (7.9) & {\bf } & \ldots & \ldots & $>$1.7 \\
18 & 38 &  12 37 41.16 &   +62 12 20.5 & 3.8 &  12 37 41.16 &   +62 12 20.9 & 3.5 & {\bf 0.032} & \ldots & 2.1 & $2.4^{+0.6}_{-2.0}$ \\
 &  &  12 37 41.63 &   +62 12 23.6 & 5.8 &  12 37 41.66 &   +62 12 23.6 & 6.0 & 0.051 & \ldots & 2.0 & $>$2.2$^c$ \\
19 & \ldots &  12 36 04.40 &   +62 07 02.7 & 2.6 &  12 36 04.43 &   +62 07 02.9 & 2.8 & {\bf 0.022} & \ldots & \ldots & $>$1.5 \\
20 & \ldots &  12 37 12.48 &   +62 10 35.4 & 3.0 &  12 37 12.51 &   +62 10 35.6 & 2.8 & {\bf 0.030} & \ldots & 3.3 & $>$2.3 \\
21 & \ldots &  12 38 00.80 &   +62 16 11.7 & 1.5 &  12 38 00.81 &   +62 16 11.7 & 1.4 & {\bf 0.016}$^a$ & \ldots & 2.3 & $>$1.6 \\
22 & \ldots &  12 36 48.60 &   +62 12 16.1 & 4.7 &  12 36 48.65 &   +62 12 15.7 & 4.2 & 0.083$^a$ & \ldots & 1.7 & $>$0.7 \\
23 & \ldots &  12 37 16.67 &   +62 17 33.2 & 1.4 &  12 37 16.67 &   +62 17 33.3 & 1.4 & {\bf 0.001} & {\bf 1.1460} & 1.6 & $>$1.8 \\
24 & 23 &  12 36 08.58 &   +62 14 35.3 & (6.4) &  12 36 08.60 &   +62 14 35.3 & (6.4) & 0.079 & \ldots & 2.7 & $2.4^{+2.0}_{-0.8}$ \\
25 & \ldots &  12 36 51.72 &   +62 05 03.0 & 4.1 &  &  &  & {\bf 0.021} & \ldots & \ldots & $>$0.7 \\
26 & 40 &  12 37 13.86 &   +62 18 26.2 & 0.6 &  12 37 13.85 &   +62 18 26.2 & 0.5 & {\bf 0.000} & \ldots & 2.6 & $>$0.7 \\
27 & \ldots &  12 37 19.62 &   +62 12 20.3 & 1.3 &  12 37 19.61 &   +62 12 20.9 & 0.9 & {\bf 0.034}$^b$ & \ldots & 3.1 & $>$0.7 \\
 &  &  12 37 20.01 &   +62 12 22.0 & 2.1 &  12 37 19.99 &   +62 12 22.6 & 2.2 & {\bf 0.050}$^b$ & {\bf 2.4600} & 2.0 & $>$0.7 \\
28 & \ldots &  12 36 44.03 &   +62 19 38.8 & 4.2 &  12 36 44.03 &   +62 19 38.4 & 3.9 & 0.071$^a$ & \ldots & 2.5 & $>$0.7 \\

\hline
\end{tabular}
\label{tab:id}
$^a$Identified in 3-$\sigma$ radio catalogue\\
$^b$Identified in MIPS 24\,\micron\ catalogue\\
$^c$Photometric redshift calculation excludes extremely noisy 850\,\micron\ photometry\\
$^d$Position for $HDF~850.1$ from \citet{dunlop2004}. See also \citet{cowie2009}.
\end{table*}

\begin{table*}
  \centering
  \caption{Photometry data listed in the same order as the
    identifications in Table~\ref{tab:id}.
    The 850\,\micron\ flux densities are
    measured directly from the SCUBA map of \citet{pope2006} at the
    positions of each proposed counterpart, and should therefore be un-biased at
    each correct position. 24\,\micron\ upper-limits are given at a 
    significance of 3-$\sigma$. De-boosted AzTEC 1.1\,mm flux densities are 
    taken from \citet{perera2008} except for $AzGN~1$, $AzGN~1.2$, $AzGN~14$ and
    $AzGN~22$, which are calculated using the raw map flux densities from 
    Table~\ref{tab:newflux}, but adopting the same prior.}
  \begin{tabular}{lrrrrrrrr}
    \hline
    AzTEC & 1.4\,GHz & 1.1\,mm & 850\,\micron\ & 24\,\micron\ & 
    8\,\micron\ & 5.8\,\micron\ & 4.5\,\micron\ & 3.6\,\micron\ \\ 
    ID & ($\mu$Jy) & (mJy) & (mJy) & ($\mu$Jy) & ($\mu$Jy) & ($\mu$Jy) & 
    ($\mu$Jy) & ($\mu$Jy) \\ 
    \hline

1 & $ 72.8 \pm 12.3$ & $11.81 \pm^{+1.18}_{-1.07}$ & $20.30 \pm 2.10$ & $  65.5 \pm  4.5$ & $ 25.3 \pm 1.5$ & $ 15.9 \pm 1.3$ & $ 9.2 \pm 0.8$ & $  6.8 \pm 0.6$ \\
1.2 & $173.5 \pm  6.3$ & $2.87 \pm^{+1.25}_{-1.25}$ & $ 9.90 \pm 2.30$ & $  20.2 \pm  3.5$ & $  9.4 \pm 1.1$ & $  6.1 \pm 1.0$ & $ 3.9 \pm 0.5$ & $  3.9 \pm 0.6$ \\
 & $ 35.2 \pm  6.3$ &   & $ 9.90 \pm 2.30$ & $  12.9 \pm  3.1$ & $ 14.9 \pm 1.5$ & $ 11.8 \pm 1.3$ & $ 9.8 \pm 0.8$ & $  8.0 \pm 0.9$ \\
2 & $ 26.2 \pm  4.6$ & $5.91 \pm^{+1.02}_{-1.00}$ & $ 5.74 \pm 2.81$ & $  37.7 \pm  4.8$ & $ 24.9 \pm 1.5$ & $ 17.0 \pm 1.3$ & $10.9 \pm 0.8$ & $  6.4 \pm 0.6$ \\
3 & $ 36.0 \pm  4.2$ & $5.35 \pm^{+0.94}_{-1.08}$ & $10.50 \pm 1.59$ & $  26.2 \pm  5.1$ & $  5.1 \pm 1.1$ & $  2.7 \pm 0.9$ & $ 2.0 \pm 0.4$ & $  1.2 \pm 0.4$ \\
4 & $ 34.1 \pm 10.7$ & $4.69 \pm^{+1.06}_{-1.06}$ & \ldots & $  92.5 \pm  4.2$ & $ 37.0 \pm 1.7$ & $ 26.0 \pm 1.5$ & $16.0 \pm 0.8$ & $  9.6 \pm 0.9$ \\
5 & $128.2 \pm  8.0$ & $4.13 \pm^{+1.08}_{-0.98}$ & $ 4.37 \pm 3.92$ & $ 181.0 \pm  6.5$ & $ 28.9 \pm 1.7$ & $ 18.9 \pm 1.3$ & $12.7 \pm 0.8$ & $  9.2 \pm 0.9$ \\
6 & $ 58.5 \pm 13.1$ & $4.13 \pm^{+1.12}_{-1.00}$ & \ldots & $  11.3 \pm  8.3$ & $  9.9 \pm 1.1$ & $  4.1 \pm 1.0$ & $ 2.5 \pm 0.4$ & $  1.7 \pm 0.4$ \\
7 & $127.3 \pm  8.6$ & $3.95 \pm^{+1.08}_{-0.98}$ & $ 6.61 \pm 1.89$ & $ 537.0 \pm  9.3$ & $ 37.8 \pm 1.7$ & $ 53.3 \pm 1.5$ & $45.0 \pm 1.0$ & $ 37.9 \pm 1.2$ \\
 & $ 51.6 \pm  8.2$ &   & $ 7.47 \pm 2.35$ & $ 219.0 \pm  6.6$ & $ 12.3 \pm 1.5$ & $ 16.1 \pm 1.3$ & $11.4 \pm 0.8$ & $  9.2 \pm 0.9$ \\
8 & $120.0 \pm  7.6$ & $3.83 \pm^{+1.08}_{-1.00}$ & $ 7.83 \pm 1.29$ & $ 145.0 \pm 12.2$ & $ 13.8 \pm 1.5$ & $ 11.2 \pm 1.3$ & $ 7.9 \pm 0.8$ & $  5.7 \pm 0.6$ \\
 & $ 94.3 \pm 11.8$ &   & $ 5.50 \pm 1.07$ & $ 422.0 \pm 29.1$ & $ 27.8 \pm 1.7$ & $ 29.8 \pm 1.5$ & $31.7 \pm 1.0$ & $ 44.6 \pm 1.2$ \\
9 & $ 26.0 \pm  5.4$ & $3.39 \pm^{+1.02}_{-1.10}$ & $ 5.65 \pm 2.38$ & $  32.2 \pm  5.0$ & $  9.2 \pm 1.1$ & $  8.5 \pm 1.0$ & $ 6.6 \pm 0.5$ & $  6.2 \pm 0.6$ \\
10 & $ 17.9 \pm  4.3$ & $3.35 \pm^{+1.02}_{-1.10}$ & $ 2.25 \pm 1.95$ & $  22.1 \pm  6.8$ & $  9.7 \pm 1.1$ & $  4.2 \pm 1.0$ & $ 2.3 \pm 0.4$ & $  1.2 \pm 0.4$ \\
11 & $ 36.2 \pm 10.2$ & $3.27 \pm^{+1.08}_{-1.08}$ & $10.94 \pm 4.60$ & \ldots & $ 22.0 \pm 2.0$ & $ 10.5 \pm 2.0$ & $ 5.6 \pm 1.5$ & $  4.6 \pm 1.5$ \\
12 & $ 26.6 \pm  5.2$ & $3.07 \pm^{+1.12}_{-1.08}$ & $ 4.42 \pm 8.50$ & $  29.7 \pm  8.8$ & $  7.9 \pm 1.1$ & $  4.7 \pm 1.0$ & $ 2.4 \pm 0.4$ & $  1.6 \pm 0.4$ \\
13 & $ 28.9 \pm 12.1$ & $3.07 \pm^{+1.10}_{-1.12}$ & $16.49 \pm 4.64$ & $ 154.0 \pm  6.3$ & $ 19.6 \pm 1.5$ & $ 21.9 \pm 1.3$ & $16.4 \pm 0.8$ & $ 12.8 \pm 0.9$ \\
14 & \ldots & $2.87 \pm^{+1.25}_{-1.25}$ & $ 5.88 \pm 0.33$ & \ldots & \ldots & \ldots & \ldots & \ldots \\
15 & $123.4 \pm  5.5$ & $3.23 \pm^{+1.26}_{-1.32}$ & \ldots & $<65.1$ & $  6.7 \pm 1.1$ & $ 14.7 \pm 1.3$ & $ 7.1 \pm 0.5$ & $  2.4 \pm 0.4$ \\
 & \ldots &   & \ldots & $<65.1$ & $ 10.9 \pm 1.5$ & $ 13.6 \pm 1.3$ & $ 7.5 \pm 0.5$ & $  2.6 \pm 0.4$ \\
16 & $ 37.8 \pm  8.2$ & $2.89 \pm^{+1.08}_{-1.14}$ & $ 5.75 \pm 0.98$ & $ 326.0 \pm  8.0$ & $ 43.4 \pm 1.7$ & $ 29.5 \pm 1.5$ & $18.1 \pm 0.8$ & $ 12.3 \pm 0.9$ \\
 & $ 29.9 \pm  9.2$ &   & $ 5.38 \pm 0.98$ & $   4.8 \pm  7.2$ & $ 27.1 \pm 1.7$ & $ 27.9 \pm 1.5$ & $19.5 \pm 0.8$ & $ 14.9 \pm 0.9$ \\
17 & $ 80.7 \pm 11.8$ & $3.23 \pm^{+1.24}_{-1.42}$ & \ldots & $<24.0$ & $ 13.3 \pm 1.5$ & $ 15.7 \pm 1.3$ & $ 8.1 \pm 0.8$ & $  2.0 \pm 0.4$ \\
 & $ 66.2 \pm  5.8$ &   & \ldots & $<59.4$ & $  6.9 \pm 1.1$ & $ 14.4 \pm 1.3$ & $10.5 \pm 0.8$ & $  8.6 \pm 0.9$ \\
 & \ldots &   & \ldots & $<58.5$ & $ 11.9 \pm 1.5$ & $ 13.5 \pm 1.3$ & $ 8.2 \pm 0.8$ & $  5.8 \pm 0.6$ \\
18 & $ 38.3 \pm  9.0$ & $2.79 \pm^{+1.16}_{-1.08}$ & $18.97 \pm 6.56$ & $ 127.0 \pm  5.8$ & $ 23.1 \pm 1.5$ & $ 14.4 \pm 1.3$ & $10.3 \pm 0.8$ & $  6.6 \pm 0.6$ \\
 & $ 32.2 \pm  5.0$ &   & $24.17 \pm 6.57$ & $ 111.0 \pm  4.6$ & $ 16.4 \pm 1.5$ & $ 11.1 \pm 1.3$ & $ 7.8 \pm 0.8$ & $  5.9 \pm 0.6$ \\
19 & $ 34.4 \pm  5.6$ & $3.07 \pm^{+1.20}_{-1.36}$ & $15.03 \pm 8.30$ & $< 9.0$ & $ 42.7 \pm 1.7$ & $ 47.0 \pm 1.5$ & $31.8 \pm 1.0$ & $ 46.6 \pm 1.2$ \\
20 & $ 26.6 \pm  4.4$ & $2.79 \pm^{+1.08}_{-1.16}$ & $ 3.98 \pm 1.53$ & $  35.1 \pm  5.7$ & $ 16.6 \pm 1.5$ & $ 14.3 \pm 1.3$ & $ 8.9 \pm 0.8$ & $  5.5 \pm 0.6$ \\
21 & $ 23.0 \pm  5.7$ & $2.65 \pm^{+1.16}_{-1.16}$ & $ 6.29 \pm 4.47$ & $ 182.0 \pm  6.2$ & $ 24.7 \pm 1.5$ & $ 37.1 \pm 1.5$ & $34.1 \pm 1.0$ & $ 26.5 \pm 1.2$ \\
22 & $ 18.9 \pm  4.3$ & $3.35 \pm^{+1.22}_{-1.02}$ & $ 0.29 \pm 0.40$ & $ 291.0 \pm  7.4$ & $ 24.0 \pm 1.5$ & $ 32.4 \pm 1.5$ & $32.0 \pm 1.0$ & $ 27.3 \pm 1.2$ \\
23 & $381.2 \pm  8.2$ & $2.39 \pm^{+1.16}_{-1.18}$ & $-4.18 \pm 3.56$ & $1240.0 \pm 15.5$ & $239.6 \pm 1.7$ & $129.3 \pm 1.5$ & $83.5 \pm 1.0$ & $ 62.7 \pm 1.2$ \\
24 & $ 44.6 \pm  8.6$ & $2.39 \pm^{+1.18}_{-1.20}$ & $ 5.54 \pm 1.74$ & $  51.1 \pm  5.8$ & $ 18.3 \pm 1.5$ & $ 13.4 \pm 1.3$ & $ 9.5 \pm 0.8$ & $  6.4 \pm 0.6$ \\
25 & $ 79.2 \pm  5.6$ & $2.55 \pm^{+1.32}_{-1.42}$ & \ldots & \ldots & \ldots & \ldots & \ldots & \ldots \\
26 & $651.8 \pm  5.0$ & $2.39 \pm^{+1.10}_{-1.28}$ & $11.53 \pm 2.43$ & $  55.2 \pm  5.8$ & $ 16.6 \pm 1.5$ & $  9.4 \pm 1.3$ & $ 6.0 \pm 0.5$ & $  3.5 \pm 0.6$ \\
27 & $ 10.1 \pm  8.1$ & $2.31 \pm^{+1.16}_{-1.22}$ & $ 0.27 \pm 2.09$ & $  31.2 \pm  6.7$ & $ 22.4 \pm 1.5$ & $ 20.2 \pm 1.3$ & $15.5 \pm 0.8$ & $ 11.7 \pm 0.9$ \\
 & $ 20.2 \pm 10.8$ &   & $ 1.18 \pm 2.19$ & $ 141.0 \pm  7.3$ & $ 17.7 \pm 1.5$ & $ 17.1 \pm 1.3$ & $12.9 \pm 0.8$ & $ 10.7 \pm 0.9$ \\
28 & $ 20.8 \pm  5.4$ & $2.31 \pm^{+1.14}_{-1.30}$ & $12.80 \pm 8.66$ & $  29.9 \pm  5.9$ & $ 10.2 \pm 1.1$ & $  4.8 \pm 1.0$ & $ 4.1 \pm 0.5$ & $  2.1 \pm 0.4$ \\

    \hline
\end{tabular}
\label{tab:sed}
\end{table*}

\section{Notes on Each Source}
\label{sec:notes}

This section gives detailed information on each source not provided in
the tables of Appendix~\ref{sec:tables}.

\subsection{Robust Identifications}
Sources in this category have potential counterparts with $P<0.05$
within 6~arcsec.

\begin{description}
\item[$AzGN~1$:] $GN~20$ from \citet{pope2006}. This source has been
  deblended from $AzGN~1.2$ (see Table~\ref{tab:newflux}), and the
  submm emission was also localized using the SMA \citep{iono2006}. A
  spectroscopic redshift of 4.055 for this source was reported in
  \citet{daddi2008} based on molecular CO emission detected with the
  IRAM PdBI., with some confirmation based on optical spectroscopy in
  Pope~et.~al~(submitted).

\item[$AzGN~1.2$:] This was already known to be a second component of
  $AzGN~1$ from the SCUBA data.  In the AzTEC map the source was
  deblended by performing a simultaneous fit of two scaled effective
  PSFs using the peak at the location of $AzGN~1$ and the position of
  $GN~20.2$ from \citet{pope2006} as starting values (see
  Table~\ref{tab:newflux}). Positions and flux densities were then
  allowed to vary (a total of 6 parameters). The reduced $\chi^2$ for
  this fit decreased to $0.99$ from $1.15$ when only a single source
  was used, justifying the addition of the extra parameters. This
  procedure yields a clear 4.2-$\sigma$ source which corresponds to
  $GN~20.2$ from \citet{pope2006}. There are two possible radio
  identifications, one fainter object is 4.1~arcsec away with
  $P=0.037$, and the other brighter source is 2.6~arcsec away with
  $P=0.005$. The brighter object is the claimed counterpart to
  $AzGN~1.2$ from \citet{pope2006}. However, both potential
  counterparts mentioned here are also discussed in
  \citet{daddi2008}. They detect a faint emission line from the
  brighter radio source which they argue to be consistent with
  molecular CO emission at a redshift 4.051.  We concur with
  \citet{daddi2008} that the brighter object is likely an AGN, based
  on its relatively large radio/mm flux ratio.

\item[$AzGN~3$:] The AzTEC detection, and proposed counterpart, are
  both coincident with the SCUBA source $GN~10$ and identification
  from \citet{pope2006}. Similar to $AzGN~1$ and $AzGN~1.2$,
  \citet{daddi2009} identified a spectroscopic redshift of 4.042 using
  CO emission detected with the IRAM PdBI.

\item[$AzGN~4$:] There is a single unambiguous radio counterpart.
 
\item[$AzGN~5$:] There is a single unambiguous radio counterpart. This
  object is not part of the \citet{pope2006} catalogue (probably
  because of the high noise in this region), but was originally
  discovered in the SCUBA jiggle map of \citet{wang2004}
  (GOODS~850--6). Our proposed counterpart corresponds to object `c'
  from their analysis. We note a large discrepancy between the
  850\,\micron\ flux density of this source at the coordinates of the
  proposed counterpart in the HDF supermap, $4.4\pm3.9$\,mJy, used in
  this paper (Table~\ref{tab:sed}), compared with $19.41\pm3.2$\,mJy
  from \citet{wang2004} which is due at least in part to the lack of
  correction for Eddington bias in that work.  Based on the 1.1\,mm
  flux density and our measured ratio $S_{850}/S_{1.1} = 1.8$, we
  would expect $S_{850}\simeq7\,$mJy.

\item[$AzGN~6$:] There is a single unambiguous radio counterpart.

\item[$AzGN~7$:] This object is also a significant SCUBA source,
  $GN~39$, described in \citet{wall2008}. We find the same two radio
  identifications with spectroscopic redshifts 1.996
  \citep{chapman2005} and 1.992 \citep{swinbank2004} respectively.

\item[$AzGN~9$:] This objects is also a significant SCUBA source,
  $GN~37$ from \citet{pope2006} for which we find the same
  identification with spectroscopic redshift 3.190 \citep{cowie2004}.

\item[$AzGN~11$:] This object is close to the SCUBA source $GN~27$
  from \citet{pope2006} that is classified as `tentative' since it
  does not exceed their threshold for posterior deboosted flux density
  probability above 0. There appear to be three distinct peaks in the
  radio map that fall between the SCUBA and AzTEC peaks (but closer to
  the AzTEC position). Two of these peaks are coincident with {\em
    Spitzer\/} sources at redshifts 0.56 and 0.276. The ACS/IRAC
  counterpart to $GN~27$ claimed in \citet{pope2006} is to the south
  of the three objects that we find, and is not a significant source
  in our 1.4\,GHz map. The third source is considerably fainter and
  redder in the IRAC channels. However, due to blending with the other
  sources we are unable to produce the SED at {\it Spitzer}
  wavelengths in Table~\ref{tab:sed}.  We rule out the low-$z$
  identifications based on their highly unlikely SEDs which would
  exhibit extraordinarily low radio/(sub)mm flux density ratios.

\item[$AzGN~12$:] There is a single unambiguous radio counterpart.

\item[$AzGN~13$:] There is a single unambiguous radio counterpart
  within 6~arcsec. We note that there are two radio sources 7.9~arcsec
  (12:35:53.90 +62:13:37.1) and 8.7~arcsec (12:35:53.24 +62:13:37.5)
  to the south that are coincident with a low-significance peak of
  SCUBA emission. The second source was also part of the
  \citet{chapman2005} sample. While they are not considered likely in
  our analysis, we list them here for completeness. Had we used a
  search radius of 10~arcsec they would have $P$ values of 0.103 and
  0.108, and lie at redshifts 0.8770 \citep{reddy2006} and 2.098
  \citep{chapman2005,reddy2006}, respectively.

\item[$AzGN~14$:] This object is $HDF~850.1$, the highest SNR SMG from
  the HDF-N map of \citet{hughes1998}. Similar to the case for
  $AzGN~1$ and $AzGN~1.2$, this source has been de-blended from
  $AzGN~22$ (see Table~\ref{tab:newflux}). Various counterparts have
  been suggested, with two detailed studies giving different answers.
  \citet{dunlop2004} subtract the emission from an elliptical galaxy
  that is believed to obscure a faint background $K$-band counterpart
  at 12:36:52.07 +62:12:25.7 (this position is given in
  Table~\ref{tab:id}). Recently, \citet{cowie2009} used the SMA to
  localize the source of the submm emission to 12:36:51.99
  +62:12:25.83, concluding that this position is incompatible with the
  \citet{dunlop2004} counterpart. For the purposes of measuring
  un-biased flux densities in the SCUBA and AzTEC maps, either
  position may be used as the differences are small compared to the
  (sub)mm beams.  Had we applied our counterpart search blindly we
  would have instead chosen a more distant radio source at 12:36:51.72
  +62:12:21.36 ($P=0.036$), although we would have rejected this
  counterpart based on the unrealistic SED. This case demonstrates the
  potential for counterparts identified with low values of $P$ to be
  misleading.  No {\em Spitzer\/} flux densities are quoted for this
  special source, and it is excluded from most of the analysis in this
  paper.

\item[$AzGN~15$:] There is a single unambiguous radio counterpart,
  although it appears to resolve into two objects in the {\em
    Spitzer\/} catalogue. Inspection of the map shows, however, that
  these data lie at the extreme edge of the {\it Spitzer} coverage,
  and we suspect that the two sets of photometry are in fact for the
  same source.

\item[$AzGN~16$:] Similar to $AzGN~7$, this object is also detected
  with SCUBA \citep[$GN~4$ from][]{pope2006}, and has two radio
  counterparts. The more distant counterpart has a spectroscopic
  redshift of 2.578 \citep{chapman2005}.

\item[$AzGN~18$:] This object is also a significant SCUBA source,
  $GN~38$ from \citet{pope2006}, and we find the same two radio
  counterparts. One of these counterparts technically misses the
  significance threshold with $P=0.051$, but lies very close to the
  centre of the SCUBA emission peak. Conversely, the other object,
  with $P=0.032$, lies closer to the peak of the AzTEC emission. The
  photometric redshifts for both sources are consistent with being at
  the same distance.

\item[$AzGN~19$:] There is a single unambiguous radio counterpart, and
  the position is coincident with a region of faint SCUBA emission. It
  falls off the edge of the MIPS 24\,\micron\ coverage, but within the
  IRAC footprint. There is a bright optical source within 1~arcsec of
  the radio position at $z=0.65$. However, we reject this redshift and
  SED combination as it would imply an extraordinarily low
  radio/(sub)mm flux ratio. Therefore the radio identification
  proposed here may be incorrect and $AzGN~19$ is associated with the
  low-$z$ optical source instead, or it is correct and unrelated to
  the optical source. Since the (sub)mm/radio photometric redshift
  estimate is consistent with a more typical SMG
  ($z_{\mathrm{rm}}>1.5$) we prefer the first hypothesis.

\item[$AzGN~20$:] There is a single unambiguous radio counterpart, and
  the position is coincident with a region of faint SCUBA emission.

\item[$AzGN~21$:] There is a single unambiguous radio counterpart.

\item[$AzGN~23$:] There is a single unambiguous radio
  counterpart. This object has the lowest spectroscopic redshift in
  the sample, $z=1.146$ \citep{cowie2004}.

\item[$AzGN~25$:] This object has a single clear radio counterpart,
  but lies beyond the {\em Spitzer\/} IRAC coverage. The MIPS
  24\,\micron\ image exhibits emission at the location of the radio
  counterpart, but is not part of the {\em Spitzer\/} catalogue
  produced by the GOODS team.

\item[$AzGN~26$:] This object is also a significant SCUBA source,
  $GN~40$ from \citet{wall2008}. There is a single unambiguous radio
  identification.

\item[$AzGN~27$:] This source is coincident with faint SCUBA
  emission. There are no radio sources with $P<0.05$, even searching
  out to 10\,arcsec. However, there are two MIPS 24\,\micron\ sources
  0.9~arcsec and 2.2~arcsec away, with $P$ values 0.034 and 0.05,
  respectively. The more distant source has a spectroscopic redshift
  $z=2.460$ (Stern et al.~in preparation), and photometric redshift
  estimates for the other source are broadly consistent with this
  value. As this object appears to have an extraordinarily faint radio
  flux density given the 24\,\micron\ measurement, we warn the reader
  that these identifications may be suspect.

\end{description}

\subsection{Other Identifications}
For the remaining sources we give our best estimates for the
counterparts, considering objects with $0.05<P<0.10$ and/or searching
out to 10~arcsec from the AzTEC positions.

\begin{description}

\item[$AzGN~2$:] There is a radio source 5.3~arcsec to the north that
  slightly misses the significance cut ($P=0.051$). This proposed
  counterpart is in the direction of a faint 2-$\sigma$ signal in the
  SCUBA map.

\item[$AzGN~8$:] This object is also a significant SCUBA source,
  $GN~12$. There are two potential radio counterparts 7~arcsec and
  9~arcsec away. Had we used a search radius of 10~arcsec, the values
  of $P$ would have been 0.037 and 0.050. Since the closer object
  would then have a lower $P$, and is also closer to the SCUBA peak,
  we prefer it as the most likely candidate, although strictly based
  on the 1.1\,mm and 1.4\,GHz emission they are both reasonable
  candidates. \citet{pope2006} were unable to obtain an optical
  spectrum for the nearer candidate. The other candidate was detected
  serendipitously with Keck at a redshift $z=2.006$. This second
  object was rejected by \citet{pope2006} based on its distance from
  the SCUBA centroid, and the unlikely rest-frame UV emission that
  would be uncharacteristic of a dusty star-forming galaxy.

\item[$AzGN~10$:] This object has a single potential radio counterpart
  within 6~arcsec, and is also coincident with faint SCUBA emission.

\item[$AzGN~17$:] This object is at the edge of the MIPS 24\,\micron\
  coverage, but within the IRAC footprint. There are two potential
  radio counterparts at distances 7.1~arcsec and 7.6~arcsec.

\item[$AzGN~22$:] This source lies on a region of extended emission to
  the south-east of $AzGN~14$, which is also seen in the SCUBA map.
  To reduce the effect of blending with $AzGN~14$ we have performed a
  simultaneous 2-object fit, as with $AzGN~1$ and $AzGN~1.2$
  (resulting positions and map flux densities given in
  Table~\ref{tab:newflux}). The most likely counterpart, given the
  resulting centroid, is a radio source 4.7~arcsec away.

\item[$AzGN~24$:] There is a single potential radio counterpart
  6.4~arcsec to the south. We consider this a likely identification
  because it falls between the AzTEC peak, and the SCUBA source
  $GN~23$ \citep{pope2006}, for which the same identification was
  proposed.

\item[$AzGN~28$:] This object has a single potential radio counterpart
  4.2~arcsec away, and is coincident with faint SCUBA emission. It has
  a redshift estimate $z_{\mathrm{ir}} = 2.5$

\end{description}

\bibliographystyle{mn2e}
\bibliography{mn-jour,refs}

\end{document}